\documentclass[12pt]{article}
\newlength{\abstractwidth}
\usepackage{epsf}

\renewcommand{\thefootnote}{\fnsymbol{footnote}}
\renewcommand{\thanks}[1]{\footnote{#1}}
\newcommand{\starttext}{
\setcounter{footnote}{0}
\renewcommand{\thefootnote}{\arabic{footnote}}}

\newcommand{\bea}{\begin{eqnarray}}
\newcommand{\eea}{\end{eqnarray}}
\newcommand{\ee}{\end{equation}}
\newcommand{\be}{\begin{equation}}

\def\cA{{\cal A}}
\def\cB{{\cal B}}
\def\cC{{\cal C}}

\def\cH{{\cal H}}

\def\cM{{\cal M}}
\def\cN{{\cal N}}
\def\cO{{\cal O}}

\def\cS{{\cal S}}

\def\cU{{\cal U}}

\def\bC{{\bf C}}

\def\bR{{\bf R}}
\def\bZ{{\bf Z}}

\def\Re{{\rm Re}}
\def\Im{{\rm Im}}

\def\half{ {1\over 2}}
\def\p{\partial}

\def\a{\alpha}
\def\b{\beta}
\def\tet{\vartheta}
\def\ep{\varepsilon}

\def\g{\gamma}

\def\g{\gamma}

\def\ch{{\rm ch\, }}

\def\cotg{{\rm cotg \, }}

\def\no{\nonumber}

\begin{document}
\starttext
\setcounter{footnote}{0}

\begin{flushright}
UCLA/07/TEP/10\\
1 May  2007
\end{flushright}

\bigskip

\begin{center}

{\Large \bf Exact half-BPS Type IIB interface solutions II: 

\medskip

Flux solutions and multi-Janus }

\vskip .7in 

{\large  Eric D'Hoker,  John Estes and  Michael Gutperle}

\vskip .2in

 \sl Department of Physics and Astronomy \\
\sl University of California, Los Angeles, CA 90095, USA

\end{center}

\vskip .5in

\begin{abstract}

Regularity  and topology conditions are imposed on the  exact  Type IIB 
solutions  on $AdS_4 \times S^2 \times S^2 \times \Sigma $ with 16 
supersymmetries, which were derived in a companion paper~\cite{degAdS4}. 
We construct an infinite class of regular solutions with varying dilaton, 
and non-zero 3-form fluxes. Our solutions may be viewed as the fully 
back-reacted geometries of $AdS_5 \times S^5$ (or more generally,
Janus) doped with D5 and/or NS5 branes. The solutions  
are parametrized by the choice of an arbitrary genus $g$ hyper-elliptic 
Riemann surface $\Sigma $ with boundary, all of whose branch points are restricted to lie on a line. For genus $0$, the Janus solution with 16 
supersymmetries and 6 real parameters is recovered; its topology 
coincides with that of $AdS_5 \times S^5$.  The genus $g\geq 1$ solutions are 
parametrized by a total of $4g+6$ real numbers, $2g-1$ of which are the 
real moduli  of~$\Sigma$. The  solutions have $2g+2$
asymptotic $AdS_5 \times S^5$ regions, $g$ three-spheres with 
RR 3-form charge, and another $g$ with NSNS 3-form charge.
Collapse of consecutive branch points of $\Sigma $ yields 
singularities which correspond to D5 and NS5 branes in the probe limit.
It is argued that the AdS/CFT dual gauge theory to each of our solutions 
consists of a $2+1$-dimensional planar interface on which terminate $2g+2$ 
half-Minkowski 3+1-dimensional space-time $\cN=4$ super-Yang-Mills theories. 
Generally, the $\cN=4$ theory in each Minkowski half-space-time may have 
an independent value of the gauge coupling, and the interface may support various operators,
whose interface couplings are further free parameters of the dual gauge theory.

\end{abstract}

\newpage

\baselineskip=16pt
\setcounter{equation}{0}
\setcounter{footnote}{0}

\section{Introduction}
\setcounter{equation}{0}

In a companion paper \cite{degAdS4}, the complete solution with 16 supersymmetries was obtained for Type IIB supergravity on 
$AdS_4 \times S^2 \times S^2 \times \Sigma$ with 
$SO(2,3)\times SO(3) \times SO(3)$ isometry.\footnote{The corresponding 
BPS equations for this geometry were obtained in \cite{Gomis:2006cu}, but no
solutions, other than $AdS_5 \times S^5$, were constructed there.}
The solutions of \cite{degAdS4} were found analytically in terms of two locally 
harmonic functions $h_1$ and $h_2$ on a Riemann surface $\Sigma$ with 
boundary. Generally, these solutions have varying dilaton and non-vanishing 
3-form RR and NSNS fluxes.
The goal of this paper is to present the construction of 
an infinite subclass of such solutions which have non-singular geometry.

\smallskip

One motivation for this investigation derives from the fact that the AdS/CFT 
duals to planar interface super-Yang-Mills theories in four dimensions are of 
this type. In particular, a planar interface theory with 16 supersymmetries 
was predicted to exist in \cite{degsusy}, and its AdS/CFT dual  was indeed 
found to be a regular supersymmetric Janus solution in \cite{degAdS4}.
(The related problem of AdS/CFT duals to defect super-Yang-Mills
theory in the probe limit was studied in \cite{DeWolfe:2001pq}.)
Another motivation stems from the similarity of the problem with the construction
of half-BPS ``bubbling geometries" in \cite{Lin:2004nb}, and of AdS/CFT duals 
to half-BPS Wilson loop operators in $\cN=4$ super Yang-Mills
in 4-dimensional space-time \cite{Lunin:2006xr,Gomis:2006sb,Yamaguchi:2006te}. 
In particular, the AdS/CFT dual to a half-BPS Wilson loop corresponds to
the geometry $AdS_2 \times S^2 \times S^4 \times \Sigma$ 
(see \cite{Lunin:2006xr});
its general solution will be obtained in a further companion paper \cite{EDJEMG3}.
A final motivation is the construction of fully back-reacted Type IIB supergravity 
solutions in which the  $AdS_5 \times S^5$ geometry is doped 
with D5 and/or NS5 branes. 

\smallskip

In the present paper, we shall derive a set of regularity and topology conditions 
under which  some of the general  solutions of \cite{degAdS4} are non-singular. 
We shall restrict attention to solutions for which one $S^2$, or the other $S^2$,
shrinks to zero size on the boundary $\p \Sigma$ of $\Sigma$. This will partition 
the boundary $\p \Sigma$ into segments, each segment corresponding to
the vanishing of either  one $S^2$, or the other $S^2$, but not both.
While the segments lie on the boundary  $\p \Sigma$, they correspond to regular
interior points of the full 10-dimensional geometry. Two consecutive segments 
meet at a point on $\p \Sigma$, which actually corresponds
to an asymptotic $AdS_5 \times S^5$ throat region. 

\smallskip

A general class of regular solutions allows for an arbitrary even number 
$2g+2$ of asymptotic $AdS_5 \times S^5$ regions, connected by a 
smooth geometry. Specifically, these solutions will be parametrized by a 
genus $g\geq 0$ hyperelliptic Riemann surface $\Sigma$ with boundary 
$\p \Sigma$, 
whose $2g+2$ branch points lie on  the real line. The harmonic functions
$h_1$ and $h_2$ will obey alternating Neumann and Dirichlet boundary
conditions on $\p \Sigma$, corresponding to whether one $S^2$ or the 
other $S^2 $ shrinks to zero radius. The regularity and topology conditions, discussed in the  preceding paragraph, will impose constraints on the zeros 
and poles of the Abelian differentials $\p h_1$ and $\p h_2$.
In particular, $\p h_1$ and $\p h_2$ will have a double pole at 
each branch point, turning the point into an asymptotic $AdS_5 \times S^5$ 
region. Regularity requires a specific ordering pattern of the real zeros of  
the Abelian  differentials $\p h_1$ and $\p h_2$ with respect to the branch points.
These conditions will be solved in this paper. The resulting solutions will be
referred to as ``multi-Janus" solutions.

\smallskip

Our half-BPS solutions have some resemblance to the ``bubbling AdS" 
solutions  found in~\cite{Lin:2004nb}.  The ``coloring" of \cite{Lin:2004nb} 
would correspond here to the alternating Neumann and Dirichlet boundary 
conditions obeyed by $h_1$ and $h_2$ on $\p \Sigma$. 
From this perspective, the solutions obtained in this paper might be referred
to as ``bubbling multi-Janus" solutions.

\smallskip

At genus $g=0$, the Janus solution with 16 supersymmetries
with 6 free parameters, first obtained in \cite{degAdS4}, is recovered. 
It indeed exhibits 2 asymptotic 
$AdS_5 \times S^5$ regions, in agreement with the above counting, 
and has the topology of  $AdS_5 \times S^5$.

\smallskip

At genus $g \geq 1$, the regular solutions are parametrized by the $2g-1$ 
real moduli of $\Sigma$, and the $2g+2$ real zeros of the Abelian differentials 
$\p h_1$ and $\p h_2$. Each solution has five further real parameters; 
one for the overall scale of the dilaton; one for the overall scale of the 
10-dimensional metric, and 3 others allowing for global $SU(1,1)$ 
S-duality rotations to the general solution with a non-zero axion field. 
The total number of free parameters 
is thus $4g+6$. The differentials $\p h_1$ and $\p h_2$ may also 
have common pairs of  complex conjugate zeros, but their positions are 
fixed in terms of the moduli and the real zeros of $\p h_1$ and $\p h_2$ 
by certain elliptic or hyperelliptic period relations, which will be presented explicitly. 

\smallskip

The genus $g\geq 1$ solutions have $2g+2$ asymptotic $AdS_5 \times S^5$ 
regions. In each of these asymptotic regions, the dilaton tends to a constant,
but the constants for different $AdS_5 \times S^5$ regions will generally be 
different from one another. 
The complete boundary of the full 10-dimensional geometry consists
of $2g+2$ four-dimensional conformal Minkowski space-times, each of which 
is obtained as the boundary in each separate $AdS_5 \times S^5$ region.
The AdS/CFT correspondence implies that a conformal field theory
will live in each of these $2g+2$ four-dimensional Minkowski space-times.
On the inside of the 10-dimensional geometry, we find $g$ homology
3-spheres which carry non-vanishing RR 3-form charge (and vanishing 
NSNS 3-form charge), and another $g$ homology spheres which carry 
non-vanishing NSNS 3-form charge (and vanishing RR 3-form charge).

\smallskip

The genus $g=1$ case will be worked out in complete detail in terms
of elliptic functions. The global geometry of the solution indeed exhibits 4 asymptotic $AdS_5 \times S^5$ regions, each with a distinct constant limiting 
value of the dilaton field, as represented in Figure 1. Existence of regular solutions 
will be shown analytically, for parameters valued in some open neighborhood.

\begin{figure}[tbph]
\begin{center}
\epsfxsize=5in
\epsfysize=3.2in
\epsffile{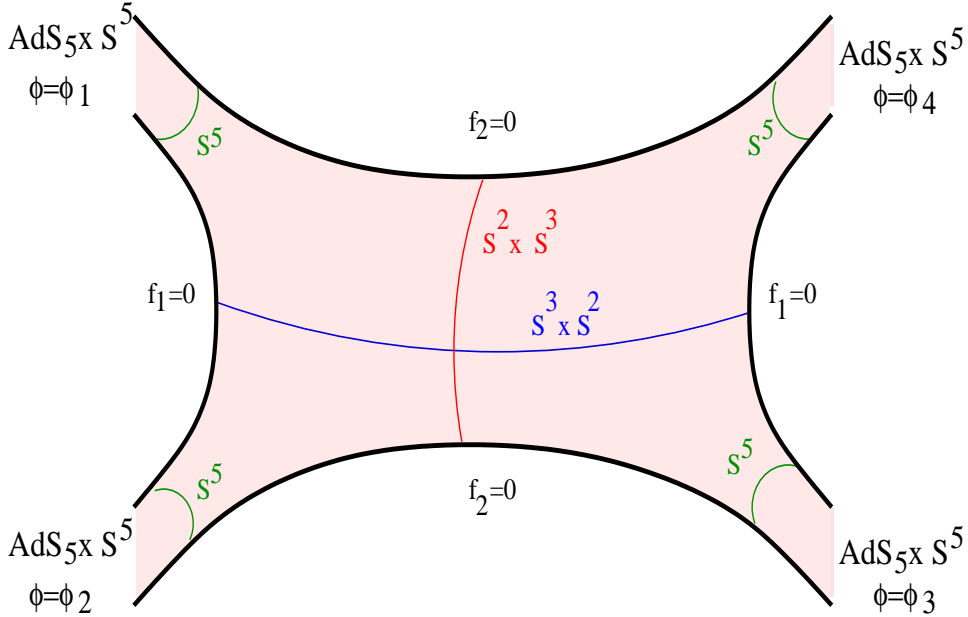}
\label{figure1}
\caption{The genus 1 solution has 4 distinct asymptotic $AdS_5 \times S^5$ regions, each with a different constant limit $\phi_1, \phi_2, \phi_3, \phi_4$ 
for the dilaton field. The radii $f_1$ and $f_2$ of the 2-spheres vanish on 
alternating segments of the boundary. The locations of the homology 3-spheres
corresponding to RR and NSNS charges are also indicated.}
\end{center}
\end{figure}

For higher genus $g$, we present an argument for the existence of 
regular solutions by induction on the genus $g$. 
(The argument applies when the free parameters are 
valued in an open neighborhood containing the  parameter space of the 
genus $g-1$ regular  solutions.) We have confirmed the existence of 
regular genus 2 solutions also by numerical analysis. 

\smallskip 

There is a sense in which the families of regular solutions for different genera 
$g$ are connected to one another. We shall show that one may pass from
a regular genus $g$ solution to a regular genus $g-1$ solution by allowing 
a branch cut between two consecutive branch points to shrink to zero size. 
The regularity and topology conditions require that the collapse of a branch 
cut on $\Sigma$ always be accompanied by the convergence to this branch 
cut of one of the complex zeros, and one real zero of either $\p h_1$ or 
$\p h_2$. A first option is to also let a second real zero converge to the  
collapsing branch cut; the regular solution at genus $g-1$ is thereby
recovered. A second option is to leave the remaining real zeros arbitrary.
The collapsing branch cut then leaves  two simple poles behind, which
are argued to correspond to a naked D5 or NS5 brane. The limit where
the residue of this pole tends to 0 yields these D5 and NS5 branes 
in the probe limit. The  solution with $g$ 
collapsed branch cuts is derived explicitly  and shown to correspond to a Janus 
geometry doped with $g$ naked D5/NS5 branes.

\smallskip

The global description of the genus $g$
parameter space of regular solutions will be presented in the form of the 
definite ordering prescription of the branch points of $\Sigma$ and the real 
zeros of the differentials $\p h_1$ and $\p h_2$, as well as the vanishing of  
the Abelian integrals of $\p h_1$ and $\p h_2$ between certain consecutive 
branch points.  The analysis of the global structure and topology of these parameter spaces poses an exciting and challenging mathematical problem, 
which is comparable to the problem of the studying the moduli space of 
instantons and magnetic monopoles.

\smallskip

The remainder of this paper is organized as follows. In section 2,
we review the Ansatz for the Type IIB supergravity fields corresponding to
the geometry $AdS_4 \times S^2 \times S^2 \times \Sigma$ 
with $SO(2,3)\times SO(3) \times SO(3)$ isometry, as well as the 
form of the complete exact solution in terms of two locally
harmonic functions $h_1$ and $h_2$ on $\Sigma$,  derived in \cite{degAdS4}.
In section 3, we derive the general regularity and topology conditions on $h_1 $, $h_2$,
and their differentials $\p h_1$ and $\p h_2$. In section 4, we introduce 
our general solution by presenting the genus $g$ hyperelliptic Ansatz, 
and subjecting it to all the regularity and topology conditions of section 3. 
These  conditions are solved in terms of a unique relative ordering of the branch 
points of $\Sigma$ and  the real zeros of the differentials $\p h_1$ and $\p h_2$.
For $g\geq 1$, the homology 3-spheres are constructed and the corresponding charges
of the RR and NSNS 3-forms are evaluated and shown to be generally
non-vanishing.
In section 5, the genus 1 case is worked out in complete detail.
In section 6, the genus 2 case is solved in part analytically, and in part numerically,
and a general argument is given for the existence of solutions to all genera.
In section 7, the collapse of branch cuts is carried out, and the presence of 
probe D5/NS5 branes is demonstrated by evaluating the 3-form fluxes on these solutions.
In section 8, a dual gauge theory is proposed as a generalization of the interface 
conformal field theory discussed in \cite{degsusy}.
In Appendix A we add poles to the differentials $\p h_1$ and $\p h_2$ that are 
on the inside of $\Sigma$, and show that these do not  lead to regular solutions.

\newpage

\section{Review of the Ansatz and the general local solution}
\setcounter{equation}{0}

In Type IIB supergravity \cite{Schwarz:1983qr,Howe:1983sr}, the bosonic 
fields  are the metric $ds^2$, the axion/dilaton fields, represented by the 
1-forms $P$ and $Q$, the complex 3-form $G$ and the self-dual 5-form $F_{(5)}$.
Upon use of the Bianchi identities, the field $P,Q$ may equivalently be 
represented by the more customary dilaton and axion fields $\Phi$ and $\chi$,
while the 3-form $G$ and the 5-form $F_{(5)}$ derive from the complex 2-form 
$B_{(2)}$ and real 4-form $C_{(4)}$ potentials. The fermionic fields are the 
gravitino $\psi _M$ and the dilatino $\lambda$. The field equations,
and BPS equations $\delta \psi _M=\delta \lambda =0$ for $\psi _M=\lambda =0$, 
as well as our conventions for Dirac matrices were presented in detail in \cite{degAdS4}. 

\smallskip

In this section, we shall review the Ansatz for the fields on the 
$AdS_4 \times S^2_1 \times S^2_2 \times \Sigma$ geometry
with $SO(2,3)\times SO(3) \times SO(3)$ isometry, as well as the 
complete solution with 16 supersymmetries derived in \cite{degAdS4}.

\subsection{The Ansatz}

The Ansatz for the metric is
\bea
\label{10dmetric}
ds^2 = f_4^2 ds^2 _{AdS_4} + f_1 ^2 ds^2 _{S_1^2} + f_2 ^2 ds^2 _{S^2_2}
+ ds^2 _\Sigma
\eea
Here, $ds^2 _{AdS_4}$, $ds^2 _{S_1^2}$ and $ds^2 _{S_2^2}$ are the 
maximally symmetric metrics respectively on $AdS_4$, $S_1^2$ and $S_2^2$ 
with unit radii;
$ds^2 _\Sigma $ is a Riemannian metric on $\Sigma$, and 
$f_1,f_2,f_4$  are real  functions on $\Sigma$. Since $\Sigma $ has a metric
and an orientation it is a Riemann surface, and we may choose local
conformal complex coordinates $w, \bar w$, in which  
$ds_\Sigma ^2  = 4 \rho ^2 |dw|^2 $, for a real $\rho$ on $\Sigma$.
It will be useful to recast the frame $e^a$, $a=8,9$ on $\Sigma $ in terms
of local complex coordinates,\footnote{We use conventions where the frame
indices are denoted by $z$ and $\bar z$,  the frame metric has non-vanishing 
components  $\delta _{z \bar z} = \delta _{\bar z  z}=2$, and the 
orientation on $\Sigma$ is given by $\ep ^{89}=+1$.}
and 
\bea
e^z = (e^8 + i e^9)/2 & = & \rho dw
\no \\
e^{\bar z} = (e^8 - i e^9)/2 & = & \rho d \bar w
\eea
The dilaton/axion fields $P$, and $Q$ are  1-forms, 
and their Ansatz is  given as follows,
\bea
P& = & p_a e^a 
\no \\
Q & = & q_a e^a
\eea
while the anti-symmetric tensor 3-form $G$ and self-dual 5-form $F_{(5)}$  are  given by
\bea
G & = & g_a e^{45a} + i h_a e^{67a}
\no \\
F_{(5)} & = & f_a (- e^{0123a} + \ep^{ab} e^{4567b} )
\eea
Here, $f_a, q_a$ are real, while $g_a, h_a, p_a$ are complex frame vectors on $\Sigma$.

\subsection{The general solution}

In \cite{degAdS4}, it was shown that the BPS equations imply certain reality 
conditions on the supersymmetry transformation spinors as well as on the components $p_a, q_a, g_a$ and $h_a$ of the Ansatz. These reality conditions allow every solution of the BPS equations to be mapped, under the 
$SU(1,1)$ symmetry of Type IIB supergravity, into a solution with vanishing 
axion field, for which $p_a, g_a$ and $h_a$ are real, and $q_a=0$. 
We shall solve the case where these conditions hold. Then applying arbitrary
$SU(1,1)$ transformations will produce all solutions
with non-vanishing axion field as well, and accounts for 3 parameters
for every solution.

\smallskip

The general solution,  with 16 supersymmetries, to Type IIB supergravity on the manifold $AdS_4 \times S^2_1 \times S^2_2 \times \Sigma$ 
with $SO(2,3)\times SO(3) \times SO(3)$ isometry, is parametrized by two 
real harmonic functions $h_1$, $h_2$ on $\Sigma$. The form $W$ of weight $(1,1)$ is ubiquitous in the solution, and 
is defined by
\bea
W  \equiv  \p_w h_1 \p_{\bar w} h_2 + \p_w h_2 \p_{\bar w} h_1 
\eea
The form $W$ is the inner product between the vector fields $\p _w h_1$ and $\p_w h_2$, 
so that $W$ will vanish when these fields are orthogonal to one another.

\smallskip

The solution for the dilaton field $\Phi = 2 \phi$ has the following form,\footnote{For notational convenience, we use $\phi = \Phi/2$ to represent the dilaton,
in accord with \cite{degAdS4}.} 
\bea
\label{dilsol}
e^{4 \phi} =
{ 2 h_1 h_2 |\p_w h_2 |^2 - h_2^2 W \over  2 h_1 h_2 |\p_w h_1 |^2 - h_1^2 W}
\eea
while the solution for the conformal factor $\rho^2$ in the $\Sigma $-metric 
is given by
\bea
\label{rhosol}
\rho^8 = { W^2 \over h_1^3 h_2^3 }
\bigg( 2 h_1  |\p_w h_2|^2 - h_2 W \bigg) \bigg( 2  h_2 |\p_w h_1|^2 - h_1 W \bigg)
\eea
The functions $f_1, f_2$, and $f_4$ entering the ten-dimensional metric
(\ref{10dmetric}) are 
most succinctly expressed by leaving some of the $\phi$- and $\rho$-dependence
manifest in the solution, 
\bea
\label{fsol}
\rho \, f_1 &=& 
	- 2 \nu\,  \Re \left ( e^{-2 \phi} |\p_w h_2|^2 - e^{2 \phi} |\p_w h_1|^2 -  i W \right )^{\half}
\no\\
\rho \, f_2 &=& 
- 2 \, \Im \, \left ( e^{-2 \phi} |\p_w h_2|^2 - e^{2 \phi} |\p_w h_1|^2 -  i W \right )^{\half}
\no \\
\rho \, f_4 &=& 
\left  | e^{-\phi} \p_w h_2  - i \, e^\phi  \p_w h_1 \right |
+ \left  | e^{-\phi} \p_w h_2  + i \, e^\phi  \p_w h_1  \right | 
\eea
Formulas direcly for the metric factors $f_1$, $f_2$ and $f_4$
were derived in Appendix E of \cite{degAdS4}. Of interest to us
in the present paper will be the following combinations, recorded
for $W\leq 0$,
\bea
\label{f1f2f4}
f_1 ^2 f_4 ^2 & = & 4 e^{+2 \phi} h_1^2
\no \\
f_2 ^2 f_4^2 & = & 4 e^{-2 \phi} h_2^2
\eea
The expression for $W>0$ will not be needed in this paper.

\smallskip

Finally, the expressions for the 3-form components $g_a$, $h_a$ 
are most usefully presented in terms of the 
$B_{(2)}$ gauge potential, as this will allow us to directly compute 3-form 
fluxes and charges. The values for the field strength $F_{(3)}=dB_{(2)}$  
may be found in
\cite{degAdS4}, but will not be needed here. We have,
\bea
\label{bsol1}
B_{(2)} = b_1 \, \hat e^{45} + i b_2 \, \hat e^{67}
\eea
Here, $\hat e^{45}$  and $\hat e^{67}$  are the volume forms for the spheres $S_1^2$ and $S^2_2$ with unit radii. The functions $b_1$ and $b_2$ are given by
\bea
\label{bsol2}
b _1 & = &  + 2 \tilde h_2 + 2 i  h_1 h_2 \,
{ \p_w h_1 \p_{\bar w} h_2 - \p_{\bar w} h_1 \p_w h_2
 \over 2 h_2 |\p_w h_1|^2 - h_1W}
\no \\
b _2 & = &  - 2 \tilde h_1 + 2 i  h_1 h_2 \,
{\p_w h_1 \p_{\bar w} h_2 - \p_{\bar w} h_1 \p_w h_2
 \over 2 h_1 |\p_w h_2|^2 - h_2 W}
\eea
where $\tilde h_1$ and $\tilde h_2$ are the real harmonic functions conjugate 
to the harmonic functions $h_1$ and $h_2$ respectively. As such, they
satisfy the conjugation relations,
\bea
\label{tildes}
\p_w h_1 & = & - i \p_w \tilde h_1 
\no \\
\p_w h_2 & = & - i \p_w \tilde h_2 
\eea
These solutions are local in the sense that the range of the local complex
coordinate $w, \bar w$, and thus the surface $\Sigma$, remains to be specified globally.

\subsection{The $AdS_5 \times S^5$ and Janus solutions}

A 2-parameter family of solutions is obtained from \cite{degAdS4}, 
\bea
\label{janus}
h_1 & = &  2 e^{- \phi_+} \Im \left ( e^w \right ) - 2 e^{- \phi_-} \Im \left ( e^{-w} \right ) 
\no \\
h_2 & = &  2 e^{+ \phi_+} \Re \left ( e^w \right ) + 2 e^{+ \phi_-} \Re \left ( e^{-w} \right ) 
\eea
on the infinite strip,
\bea
\label{janusdomain}
\Sigma = \left \{ w \in \bC; \, 0 \leq \Im (w) \leq {\pi \over 2} \right \}
\eea
The dilaton $\phi$ tends to the constants $ \phi_\pm $ as $\Re(w) \to \pm \infty$. When $\phi _+= \phi_-$, this is just the $AdS_5 \times S^5$ solution, while for $\phi_+ \not= \phi_-$, it is the Janus solution with 16 supersymmetries,
predicted in \cite{degsusy} on the basis of its AdS/CFT dual interface super 
Yang-Mills theory, and calculated explicitly, including the expressions of 
$\phi, \rho, f_1, f_2, f_4, g_a, h_a, f_a$, in \cite{degAdS4}. 

\newpage

\section{Regularity and topology conditions}
\setcounter{equation}{0}

To extend the general local solution, given in terms of  harmonic functions 
$h_1$ and $h_2$, to a global non-singular solution, we need to specify  further
global conformal data, namely the domain $\Sigma$ on which $h_1$ and $h_2$ 
are  harmonic. Choosing generic $\Sigma$, $h_1$, and $h_2$ will generally lead
to singularities in the dilaton $\phi$ and other functions which specify the Ansatz,
and thus to singular solutions to Type IIB supergravity. Regularity
will require interrelations between $\Sigma$, $h_1$ and $h_2$, which we shall
now exhibit.

\subsection{Basic regularity conditions}

We shall adopt the following general regularity conditions on all the 
solutions considered in this paper. The dilaton $\phi$ and the metric functions 
$\rho ^2$, $f_1$, $f_2$, and  $f_4$ are 
\begin{description}
\item[(R1)]  {\sl non-singular in the interior of $\Sigma$};
\item[(R2)] {\sl non-singular on the boundary $\p \Sigma$, 
except possibly at isolated points.} 
\end{description}
The $AdS_5 \times S^5$ asymptotic region is one example of an isolated
point in $\p \Sigma$ where $f_4$ diverges. But the nature of this singularity 
on $\p \Sigma$ is well-understood; it corresponds to a regular 10-dimensional 
geometry, and thus must be allowed. The  regularity conditions (R1) and (R2) 
will also allow for the singularities of  probe D5 and/or NS5 branes, 
for which $f_1$, $f_2$, and $f_4$ all diverge at  isolated points on $\Sigma$. 
The fact that local solutions which have poles in the interior of $\Sigma$ lead 
to singular geometries is demonstrated for a generic class of singularities in 
Appendix~A.

\subsection{Topology conditions}

In this paper, we shall restrict attention to solutions whose boundary
is locally that of $AdS_5 \times S^5$. Poles in $\p h_1$ and $\p h_2$,
located at isolated points on the boundary $\p \Sigma$ will correspond to
asymptotic throat regions, as in the $AdS_5 \times S^5$
and supersymmetric Janus solutions. The remaining part of the boundary
consists of open segments. We shall require that these open segments
of $\p \Sigma$ correspond to regular interior points (and not to boundary points)
of the full 10-dimensional  Type IIB solution. This can be achieved
by requiring that throughout each such segment, one $S^2$, or the other
$S^2$, but not both, shrink to zero size. This topological condition already
holds on the $AdS_5 \times S^5$ and Janus solutions, where it
guarantees that the  spheres $S_1^2$ and $S_2^2$ shrink to
zero size in a manner precisely needed to recover the overall $S^5$ topology. 

\smallskip

To investigate this topology, it is useful to consider the following combinations 
of the radii $f_1$ and $f_2$, readily derived from (\ref{rhosol}),
\bea
\label{f12sol}
\rho ^4 f_1^2 f_2^2 = 4 W^2
\eea
As a result of our earlier topology condition, the product $f_1 f_2$ must vanish along the entire boundary $\p \Sigma $. Using the regularity condition (R2) 
that $\rho^2$ is non-singular on $\p \Sigma$, except possibly at isolated points, 
as well as equations  (\ref{f12sol}), 
it is clear that the vanishing of $f_1 f_2$ on $\p \Sigma$ implies, 
\bea
\label{W0}
W(w, \bar w) =0 \qquad {\rm for ~ all} \qquad w \in \p \Sigma
\eea 
For the Janus solution of (\ref{janus}), $\Sigma$ of (\ref{janusdomain}) 
consist  of a strip $w \in \bR \, \oplus \,  i [0,\pi/2]$, and $f_1$ vanishes at $0$ 
while $f_2$ vanishes at $\pi/2$. Furthermore, we have  
$W =  - 4 \, \ch (\phi_+ - \phi_-) \sin (2 \, \Im (w))$, which indeed 
vanishes throughout the boundary $\p \Sigma$ of $\Sigma$ in (\ref{janusdomain}).

\smallskip

We shall also require that $\p \Sigma$ have only a  single connected component. 
From this condition, it follows that $f_1 f_2$ and $W$ cannot 
vanish in the interior of $\Sigma$, except at isolated points. By continuity, 
the sign of  $W$ must remain constant throughout the interior of $\Sigma$. 
If it were not, there would arise a curve on the inside of $\Sigma$ where $W=0$, 
which would contradict the requirement of a single boundary component. 
Since we seek families of solutions connected to $AdS_5\times S^5$, 
for which $W< 0$ inside $\Sigma$, we shall require $W<0$ on the inside of $\Sigma$, except at isolated points where $W$ may vanish.

\subsection{Interior zeros}

Negativity of $W$ greatly restricts the allowed zeros of the Abelian 
differentials $\p h_1$ and $\p h_2$. We shall now argue that all 
zeros of $\p h_1$ and $\p h_2$ in the interior of $\Sigma$ must be common.
For example, if $\p h_1$ has a zero of order $m$ at an interior point $w_0 \in \Sigma$,
then $\p h _2$ must have a zero of order precisely $m$ at $w_0$ as well.

\smallskip

To show this, choose local coordinates $v , \bar v$ around $w_0$ so that 
$v=0$ at $w_0$, and 
\bea
\p h_1 & = & \p _v h_1 dv = v^{m_1} dv + \cO(v^{{m_1}+1}dv) 
\no \\
\p h _2 & = & \p_v h_2 dv = c v^{m_2}dv  + \cO(v^{m_2+1}dv)
\hskip 0.7in c\not= 0
\eea
Parametrizing $v$ and $c$ by polar coordinates $v = |v| e^{i \theta}$, 
and  $c = |c| e^{i \gamma}$, we readily evaluate  
\bea
W = 2 |c| |v|^{m_1 + m_2} \cos \bigg ( (m_1-m_2) \theta - \gamma \bigg )  
+ \cO\left ( |v|^{m_1+m_2+1}  \right )
\eea
Since $w_0$ is an interior point, the range of $\theta$ is over the full circle,
$0 \leq \theta \leq 2\pi$. The only manner in which we can have $W<0$ for 
sufficiently small $|v|$ and all $\theta \in [0, 2 \pi]$ is if $m_1=m_2$, which proves
our earlier assertion. It then suffices to require $\cos \gamma  <0$.

\subsection{Boundary conditions on $h_1$ and $h_2$}

Using (\ref{f1f2f4}), sharp boundary conditions on $h_1$ and $h_2$ may  
be obtained. First, we argue that the $AdS_4$-radius $f_4$ cannot vanish.
If follows from equation (6.26) of \cite{degAdS4} that, if $f_4$ vanished, then 
we also must have $f_1=f_2=0$, resulting in an unphysical singularity of the 
10-dimensional Type IIB geometry. In particular, in the $AdS_5\times S^5$ 
and Janus solutions, the function $f_4$ remains bounded away from 0.
Combining the facts that $f_4 \not= 0$, and that $\phi$ is non-singular (except perhaps  at isolated points) with equations (\ref{f1f2f4}), it is clear that $f_1$ vanishes if and only if  $h_1=0$, while $f_2$ vanishes if and only if $h_2=0$. 

\smallskip

Next, we shall show that, {\sl if $h_2=0$ in an open neighborhood 
$\cU_0 \subset \p \Sigma$, then $h_1 \not=0$ in $\cU_0$, 
except possibly at isolated points, and $h_1$ satisfies Neumann boundary 
conditions in $\cU_0$.} 
The same statement, but with the roles of $h_1$ and $h_2$ interchanged, 
also holds. 

\smallskip

To prove that $h_1$ satisfies Neumann boundary conditions throughout 
$\cU_0$, we use the fact that $h_2$ is harmonic in $\Sigma$, and 
vanishes on $\cU_0$, to choose 
conformal coordinates $w=x+iy$ (with $x,y$ real), such that $h_2 = y$ in 
an open neighborhood $\cS_0 \subset \Sigma$ which contains  $\cU_0$. 
In terms of this coordinate $w$ on $\cS_0$, $W$ takes the following form,
\bea
W = - \Im (\p _w h_1) = - \p_y h_1
\eea
The  vanishing of $W$ on the boundary $\p \Sigma$, as required by (\ref{W0}),
implies that $\p_y h_1=0$ for all points in $\cU_0$,
i.e. $h_1$ satisfies Neumann boundary conditions on $\cU_0$. 

\smallskip

To prove that $h_1\not= 0$ in $\cU_0$, except possibly at isolated 
points, we make use of the expression for the dilaton field (\ref{dilsol})
on our solutions. Each term in the numerator and denominator of (\ref{dilsol})
vanishes on $\cU_0$, either because $W=0$, or because $h_2=0$, or both.
Since with our choice of coordinates $w$ we have $\p_w h_2 = -i/2$, 
which is non-vanishing, the first term in the numerator
of (\ref{dilsol}) dominates the second. Neglecting the second term allows us
to simplify by a factor of $h_1$ and we are left with,
\bea
e^{4 \phi} \sim { 1 \over 4 |\p_w h_1|^2 + 2y^{-1} h_1 \p_y h_1}
\eea
If $h_1$ were to vanish throughout an open set $\cU_0' \subset \cU_0$,
then $|\p_w h_1|^2 $ must also vanish there, since then we would have both 
$\p_x h_1=0$ and $\p_y h_1=0$. Furthermore, since $\p_y h_1=0$ in $\cU_0'$,
it follows that, if $h_1$ also vanishes there, we actually must have 
$y^{-1} h_1 \to 0$ as $y\to 0$. As a result,  the denominator has to vanish 
as $y\to 0$, and the dilaton would be singular throughout the open set 
$\cU_0'$, which contradicts assumption (R2) that the dilaton must be 
non-singular  on $\p \Sigma$, except possibly at isolated points.
Thus, we must have $h_1\not=0$ on $\cU_0$, except possibly at 
isolated points. This completes the proof of our above statement.

\smallskip

In summary, we have a remarkable conclusion.
The boundary $\p \Sigma$ is partitioned into two open sets 
$\p \Sigma _+$ and $\p \Sigma _-$. The closures of $\p \Sigma _+$
and $\p \Sigma _-$  intersect at isolated points, and their union is $\p \Sigma$.
The following  boundary conditions on the functions $h_1$ and $h_2$ now hold,
\bea
\label{DN}
{\rm on} ~ \p \Sigma _+ & \hskip 1in & f_1=h_1=0 \hskip 0.5in h_1= {\rm vanishing ~ Dirichlet}
\no \\ && \p _n h_2=0 \hskip 0.7in h_2 = {\rm Neumann}
\no \\ && \no \\
{\rm on} ~ \p \Sigma _- & \hskip 1in & f_2=h_2=0 \hskip 0.5in h_2= {\rm vanishing ~ Dirichlet}
\no \\ && \p _n h_1=0 \hskip 0.7in h_1 = {\rm Neumann}
\eea
where $\p_n $ denotes the derivative normal to the boundary $\p \Sigma$.

\subsection{Summary of all regularity and topology conditions}
\label{sectionreg}

Having analyzed the behavior of the harmonic functions $h_1$ and $h_2$
on the boundary $\p \Sigma$, it remains to determine their allowed behavior 
on the inside of $\Sigma$. Given that $f_4$ cannot vanish in $\Sigma$, 
as argued in the first paragraph of subsection 3.4, it follows from (\ref{f1f2f4}) 
that $h_1$ and $h_2$ cannot vanish on the inside of $\Sigma$.
Indeed a line or domain of zeros would force $f_1$ or $f_2$ to vanish in the 
interior of $\Sigma$, which is inconsistent with the assumption of subsection 3.2 
that $\p \Sigma$ consists of only a single connected component. For the Janus solution in (\ref{janus}), we have $h_1, h_2 >0$ in the interior of $\Sigma$. 
Since our solutions will be  connected to $AdS_5 \times S^5$,
we require $h_1, h_2 >0$ on the inside of $\Sigma$ throughout.

\medskip

In summary, we have the following regularity conditions, in addition to (R1) and (R2),
\begin{description}
\item[(R3)] On $\p \Sigma _+$,  $h_1$ and $h_2$ obey respectively 
Dirichlet and Neumann boundary conditions;\\
On $\p \Sigma _-$,  $h_1$ and $h_2$ obey respectively 
Neumann and Dirichlet boundary conditions;\\
Note that these two  conditions together  imply $W=0$ throughout $\p \Sigma$.
\item[(R4)] All zeros of $\p h_1$ and $\p h_2$ on the inside of $\Sigma$ must be common;
\item[(R5)] $W<0$ on the  inside of $\Sigma$, except possibly at isolated points, where $W=0$;
\item[(R6)] $h_1>0$ and $h_2 >0$ on the inside of $\Sigma$;
\item[(R7)] All Dirichlet boundary conditions must be vanishing, as given in (\ref{DN}). 
\end{description}

Remarkably, this combination of regularity and topology assumptions,  
leads to boundary conditions akin to those of 2-dimensional 
electro-statics, with Dirichlet and Neumann components corresponding
respectively to perfect conductor and perfect insulator.

\newpage

\section{The hyperelliptic Ansatz}\label{secfour}
\setcounter{equation}{0}

We shall now present a construction for the harmonic functions $h_1$ and 
$h_2$,  in terms of hyperelliptic Riemann surfaces of genus $g$, which 
automatically solve the conditions (R1-R5) of section \ref{sectionreg}.
Conditions (R6) and (R7) are more subtle, and will lead to certain relations on
the parameters, which we shall derive explicitly. The genus $0$ case reduces 
to the  Janus and $AdS_5 \times S^5$ solutions, which are completely 
non-singular. 
For general genus $g$, the Ansatz has $4g+6$ parameters 
(including one for the overall shift of the dilaton one for the overall scale 
of the 10-dimensional metric, and 3 for the $SU(1,1)$ rotation parameters 
to the general solution with non-vanishing axion.) In the next section, all regularity 
conditions will be solved analytically  for the case of genus 1, 
and we shall prove that, in a
certain range of these parameters, the full geometry of the solution is non-singular.

\subsection{The Janus solution re-expressed on the lower half-plane}

To begin construction of the hyperelliptic Ansatz, we map the Janus and 
$AdS_5 \times S^5$ solutions to the lower half-plane  $\Sigma$, on which 
the complex coordinates will be denoted by $u, \bar u$ with $\Im (u)\leq 0$. 
This may be done  with the help of the exponential  mapping,
\bea
- \sqrt{u} = e^{- w - (\phi_+ - \phi_-)/2}  \hskip 1in  0 \leq \Im (w) \leq {\pi \over 2}
\eea
where $\Re(w), \Re(u)$ take values through ${\bf R}$. Notice that $\Re (w) = - \infty$
and $\Re(w) = + \infty$ are respectively mapped to $u=\infty$ and $u=0$.
The harmonic functions $h_1,h_2$ and their differentials $\p h_1, \p h_2$ are given by,
\bea 
\label{janushalf}
h_1(u) =  i \, {r-u \over \sqrt{u}} + c.c.
& \hskip 1in & 
\p h_1  = - {i\over 2}   \, {u +r \over  \sqrt{u}^3} \, du
\no \\
h_2(u) =  - {1+u \over \sqrt{u}} + c.c.
& \hskip 1in &
\p h_2  = - {1 \over 2} \, {u -1 \over  \sqrt{u}^3} \, du
\eea
Here, we use the shorthand $r= e^{2 \phi_- - 2 \phi_+}$ and, to simplify the 
form of the differentials, we have omitted overall multiplicative constants in 
$h_1$ and $h_2$ whose effect is to shift the dilaton by an overall constant;
these factors may be easily restored.
From the differentials, we see that $\p h_1$ is real along the negative real
axis and imaginary along the positive real axis, and thus respectively obeys Neumann
and Dirichlet boundary conditions along these segments. For $\p h_2$, the situation is
reversed, as depicted in Figure 2. Thus, the solution (\ref{janushalf}) 
satisfies the conditions (\ref{DN}) with $\p \Sigma _+ = ]0, +\infty[$ and
$\p \Sigma _- = ]-\infty, 0[$. It was shown in \cite{degAdS4} that the solution
is everywhere non-singular, so that all conditions (R1-R7) are in fact satisfied.

\begin{figure}[tbph]
\begin{center}
\epsfxsize=3.5in
\epsfysize=2.0in
\epsffile{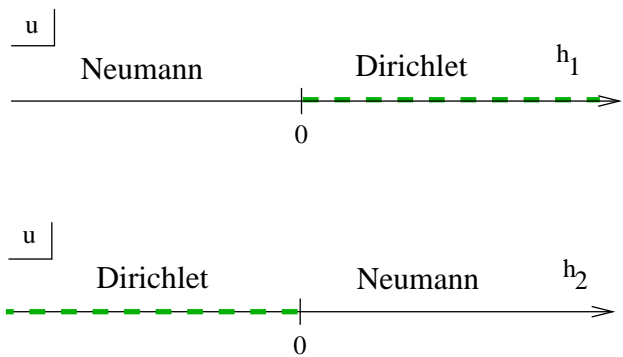}
\label{figure2}
\caption{The cut plane for $h_1$ and $h_2$ of the Janus solution.}
\end{center}
\end{figure}

\subsection{The hyperelliptic Ansatz}

The Janus solution may be generalized by having, instead of just a single
vanishing Dirichlet~(D) segment $]0, +\infty[$, and a single Neumann~(N) segment 
$]-\infty, 0[$  for $h_1$, and opposite boundary conditions for $h_2$, 
a larger number of such segments. 
This may be achieved by arranging the differentials $\p h_1$ and $\p h_2$
to have alternating real and purely imaginary values over several
segments along  $\p \Sigma$. Since topology requires the boundary 
$\p \Sigma$ to have  only a single connected component, we may 
conformally map  $\p \Sigma$ to the entire real axis, and the interior of 
$\Sigma$ to the  lower half-plane. The hyperelliptic Ansatz is obtained 
by having $2g+2$  instead of 2 separate segments
of alternating D and N boundary conditions on the real axis $\p \Sigma$.

\smallskip

Multiple alternating D and N boundary conditions may be obtained  in terms of 
$2g+2$ branch points on the real line, $e_1, e_2, \cdots , e_{2g+1}$, and 
$e_{2g+2}=\infty$, and the polynomial
\bea
\label{seq}
s^2 = (u-e_1) \prod _{k=1} ^g (u-e_{2k}) (u-e_{2k+1})
\eea
Here, we have used the $SL(2,\bR)$ symmetry on the lower  complex half-plane 
to fix one branch point at $\infty$; two further branch points may  be chosen at arbitrary 
points as well, using $SL(2,\bR)$. The algebraic equation (\ref{seq}) defines a 
hyperelliptic surface of genus $g$. Its $2g-1$ real moduli for $g\geq 1$, 
may be parametrized by the $2g-1$ remaining  branch points. 
It will be convenient  to prescribe the following definite ordering for the branch points, 
\bea
e_{2g+1} < e_{2g} < e_{2g-1} < \cdots < e_3 < e_2 < e_1
\eea
The collapse of any pair of consecutive branch points will produce a non-separating degeneration to a hyperelliptic surface of genus $g-1$.

\smallskip

To produce a suitable generalization which satisfies the regularity conditions 
(R1) and (R2), we shall assume that  the singularities of the differentials  
$\p h_1$ and $\p h_2$ at each branch point are no worse than those of the 
$AdS_5 \times S^5$ solution, i.e. of the form  $du / (u-e_i)^{3/2}$.
The differentials then have at most double poles at the branch points.
A double pole at $u=\infty$ translates to the asymptotic behavior $du/\sqrt{u}$ 
as $u \to \infty$. Poles on
the inside of $\Sigma$ would lead to singular solutions, as shown in Appendix A,
and are thus excluded by (R1) and (R2).\footnote{Certain types of poles on the real
axis will lead to mildly singular space-time solutions, which are of the 
type of D5 or NS5 branes in the probe limit. These solutions will in fact be
recovered as limits of regular hyperelliptic solutions, as will be demonstrated 
in section 7.}
Combining these requirements, we have the following form of the differentials,
\bea
\p  h_1  & = & -i \, { P_1(u) \over s(u)^3} du
\no \\
\p  h_2  & = & - { P_2(u) \over s(u)^3} du
\eea
where $P_1(u)$ and $P_2(u)$ are polynomials of degree $3g+1$ in $u$.
The overall $-$ sign for both $\p h_1$ and $\p h_2$ has been introduced 
later convenience. 

\smallskip

To realize condition (R3)  on the differentials, $\p h_1$ and $\p h_2$ need to 
alternate between taking  real and purely imaginary values on the real axis.
Along the real axis, the sign of $s^2 $ behaves as follows,
\bea
s^2 >0 & & 
u \in  \p \Sigma _+ \equiv ~ ]e_1, + \infty[ \, \cup \, 
\bigcup _{i=1} ^g ] e_{2i +1}, e_{2i}[  
\no \\
s^2 <0 & & 
u \in \p \Sigma _- \equiv~  ]-\infty , e_{2g+1} [ \, \cup \, 
\bigcup _{i=1} ^g ] e_{2i }, e_{2i-1} [
\eea
so that the denominator $s(u)^3$ alternates between being real and purely
imaginary on the real axis. Thus, to realize (R3), we must require that $P_1(u)$
and $P_2(u)$  either both be real, or both be purely
imaginary on the real axis. Without loss of generality, we may choose 
both $P_1$ and $P_2$ to be real, i.e. to be polynomials with all real coefficients.
The boundary conditions (R3) are then automatically satisfied, and we have
\bea
u \in \p \Sigma _+ & \hskip 0.5in & \p_u h_1 = \, {\rm imaginary}  \hskip 0.55in {\rm Dirichlet}
\no \\
& \hskip 0.5in & \p_u h_2 = \, {\rm real}  \hskip 1in {\rm Neumann}
\no \\
u \in \p \Sigma _- & \hskip 0.5in & \p_u h_1 = \, {\rm real}  \hskip 1in {\rm Neumann}
\no \\
& \hskip 0.5in & \p_u h_2 = \, {\rm imaginary}  \hskip 0.55in {\rm Dirichlet}
\eea
Note that, given the alternating structure of the D and N boundary conditions, 
it would have been unnatural for our solutions to use either $h_1$ or $h_2$
as the real or imaginary part of a global system of conformal coordinates.
on $\Sigma$.

\subsection{Structure of the complex zeros}

Condition (R4) of subsection \ref{sectionreg} requires all the interior zeros, 
for which $\Im (u) < 0$,  of $\p h_1$ and $\p h_2$ to be common. 
The polynomials 
$P_1$ and $P_2$ have real coefficients, so that their zeros are either real or come 
in complex pairs. These complex zeros must be common between $P_1$ and $P_2$ 
in view of (R4), so that we have the decomposition, 
$P_1(u) = P(u) Q_1(u)$ and $P_2(u) = P(u) Q_1(u)$, where $P(u)$ is a 
real polynomial whose roots are all complex and $Q_1(u)$ and $Q_2(u)$ 
are polynomials with only real roots. Thus, we obtain the more specific forms,  
\bea
\label{diffh1h2}
\p  h_1  & = & - i \, { P(u) Q_1(u)\over s(u)^3} du
\no \\
\p  h_2  & = &  - { P(u) Q_2(u) \over s(u)^3} du
\eea
We shall assume that the roots of $Q_1$ are distinct; coincident roots may be attained 
as a limit later. 
We shall parametrize and order  the roots of these polynomials as follows,
\bea
\label{PQ1Q2}
P(u) & = & \prod _{a=1}^p (u- u_a) ( u - \bar u_a) \hskip 1in \Im (u_a) < 0
\no \\
Q_1(u) & = & \prod _{b=1}^q (u- \a _b) \hskip 1.3in \a_q < \a_{q-1} < \cdots < \a_2 < \a_1
\no \\
Q_2 (u) & = & \prod _{b=1}^q (u- \b _b) \hskip 1.3in \b_q <  \b_{q-1} <  \cdots < \b _2 < \b_1
\eea
Here, $\a_b, \b _b \in {\bf R}$ and $3g+1 = 2p + q$ in view of the fact that  
$P_1$ and $P_2$ have degree $3g+1$.

\subsection{Negativity of $W$}

Condition (R5) of subsection \ref{sectionreg} requires that  $W<0$ for all 
$\Im (u)< 0$, except possibly at isolated points, where we may have $W=0$.
Clearly, at the common complex zeros  $u_a$ both differentials 
$\p h_1$ and $\p h_2$ and thus $W$ all vanish. The points $u_a$ are, of course, 
isolated points, and these zeros of $W$ fit with the terms of condition (R5).
To investigate the condition $W<0$ elsewhere, we 
recast $W$  in terms of the polynomials $P, Q_1, Q_2$, 
\bea
W = - { |P(u) Q_1(u)  |^2 \over |s|^6} \left (  i {Q_2(u) \over Q_1(u)}
- i {Q_2 (\bar u) \over Q_1(\bar u)} \right )
\eea
Since all the zeros of $Q_1$ are real and simple, and $Q_1$ and $Q_2$ are of the
same degree $q$ and start with the same highest monomial $u^q$, the ratio $Q_2/Q_1$
may be decomposed as follows,
\bea
\label{poledecomp}
Q(u) \equiv {Q_2 (u) \over Q_1(u)} = 
1 - \sum _{b=1} ^q { \g _b \over u - \a _b}
\eea
where the residues $\g_b$ are real. As a result, $W$ takes the form,
\bea
W = -i (u- \bar u) { |P(u) Q_1(u)  |^2 \over |s|^6}
\sum _{b=1} ^q { \g _b \over |u - \a _b|^2}
\eea
The requirement $W\leq 0$ in the lower half-plane is equivalent to 
\bea
\g _b \geq 0 \hskip 1in b=1, \cdots , q
\eea
Notice that if $\g_b =0$ then $\a _b$ is a zero common to $Q_1$ and $Q_2$.
We begin by assuming that $\g_b >0$ for all $b=1, \cdots , q$, the alternative 
may be attained as a limiting case hereof. The condition $\gamma _b >0$ is equivalent 
to an ordering condition between the zeros of $Q_1$ and $Q_2$, as shown 
in the Lemma below.

\smallskip

{\bf Lemma} 

\smallskip

\noindent
 {\sl (a) For $\g _b >0$, $b=1, \cdots , q$, the zeros $\beta _b$ of $Q_2$ and the 
 zeros $\a_b$  of $Q_1$, alternate, 
\bea
\label{alter}
 \a_q < \b _q < \a_{q-1} < \b_{q-1} < \cdots < \a_2 < \b _2 < \a _1 < \b _1
\eea
(b) Vice-versa, if $Q_1(x)$ and $Q_2(x)$ have highest monomial $x^q$ and 
real zeros which alternate as in (\ref{alter}), then $Q(x)=Q_2(x)/Q_1(x)$ admits 
a pole decomposition (\ref{poledecomp}) with $\g_b >0$.}

\medskip

To prove part (a), we analyze the behavior of $Q(x)$ for real $x$. Since 
all $\g _b >0$, none of the $\a _b$ is a zero of $Q_2$ or $Q$. On the other hand, 
we have, 
\bea
Q'(x) = \sum _{b=1} ^q {\g_b \over (x-\a _b)^2} >0
\eea
so that $Q(x)$ is monotonically increasing between any two consecutive poles.
Indeed, the behavior close to a pole $\a_b$ is 
\bea
Q(x) \sim { -\g_b \over x- \a_b}
\eea
so that $Q(x)$ tends to $+\infty$ on the left and to $-\infty$ on the right of every pole.
Thus, within an interval $[\a _{b+1}, \a_b]$ between any two consecutive poles,
$Q(x)$ goes from $- \infty$ (at $\a_{b+1}$) to $+\infty$ (at $\a _b$). Therefore, 
it must attain the value zero once and only once in that interval. Therefore, 
in the interval $[\a _q , \a _1]$ the function $Q_2(x)$ has precisely $q-1$  zeros.
The single remaining zero of $Q_2(x)$ is on the right
of this interval, which proves part (a) of the Lemma.

\medskip

To prove part (b) of the Lemma, we simply notice that 
\bea
\g_b = - {Q_2 (\a_b) \over Q_1 '(\a _b)} 
= - (\a_b - \b _b) \prod _{a\not=b} {(\a _b - \b _a) \over (\a _b - \a _a)}
\eea
By inspection of (\ref{alter}), it is clear that, given $b$, there are just as many 
$\a_a > \a_b$ as there are $\b _a >\a_b$, so that the product of ratios in the 
above formula is always positive. Since we also have $-(\a_b - \b _b) >0$, it 
follows that $\g_b >0$, and this completes the proof of the Lemma.

\medskip

We conclude that the condition $W<0$ throughout the lower half-plane, for polynomials
$Q_1$ with zeros $\a_1 , \cdots , \a_q$, and $Q_2$, with zeros $\b_1 , \cdots , \b_q$, 
 is equivalent to the alternation of their roots in (\ref{alter}), which thus gives a complete description and parametrization of $W<0$. The limit of coincident
zeros $\a_c = \a _{c+1}$ for some $1\leq c \leq q-1$ forces $\b _c = \a_c$.

\subsection{Obtaining the harmonic functions $h_1$, $h_2$}
\label{sectionh1h2}

Thus far, our attention has focussed on the differentials $\p h_1$ and $\p h_2$.
The full supergravity solution, of course, also involves the harmonic functions 
$h_1$ and $h_2$ themselves, as will the final two regularity conditions 
(R6) and (R7) of section subsection \ref{sectionreg}. 
From the differentials $\p h_1$ and $\p h_2$,
the harmonic functions are obtained as hyperelliptic  Abelian integrals with 
double poles at each branch point. It will be important to simplify these  
Abelian integrals and express them in terms of Abelian integrals with a pole
only at infinity. This is carried out in this subsection.

\smallskip

The functions $h_1$ and $h_2$ will have simple poles at the branch points. 
It will be convenient to expose these poles before analyzing the boundary conditions 
on $h_1$ and $h_2$. A convenient basis for meromorphic functions with a single pole 
at branch point $e_i$ is given by
\bea
{s(u) \over u- e_i}
\eea
A useful formula is obtained by taking the differential of these basis functions.
Recall that the differential $du/ s(u)$ is holomorphic, with a single zero at $\infty$,
which is of order $2g-2$.
To take the differentials of the functions $s(u)/(u-e_i)$, we introduce the following notation, 
\bea
\label{sdec}
s(u)^2 & = & (u-e_i) F_i (u)
\no \\
F_i(u) & = & (u-e_i) \bigg  ( G_i (u) + F_i'(u) \bigg ) + E_i
\no \\
E_i & \equiv & F_i (e_i)
\eea
By construction, we have $G_i(e_i)=0$.
The polynomials $s(u)^2, F_i(u), G_i(u)$ are of respective degrees $2g+1, 2g$, 
and $2g-1$, while $E_i$  do not depend on $u$. Using (\ref{sdec}), we may recast 
the double poles in terms of total derivatives of simple poles, so we shall use this 
formula in the following manner,
\bea
\label{doublepoles}
{E_i \, du \over (u-e_i) s(u) } =  - G_i(u)  {du \over s(u)}
- 2 d \left ( { s(u) \over u-e_i} \right )
\eea
To apply this formula to the differentials  $\p h_1$, $\p h_2$, 
we decompose the ratios $P(u) Q_{1,2}(u)/s(u)^2$ 
onto the double poles at the branch points, and recast the differentials
as follows,
\bea
\p h_1 & = &  -i \left (
R_1(u) - \sum _{i=1} ^{2g+1} { E_i A_i \over u - e_i} \right ) {du \over s(u)}
\no \\
\p h_2 & = &  - \left (
R_2(u) - \sum _{i=1} ^{2g+1} { E_i B_i \over u - e_i} \right ) {du \over s(u)}
\eea
where $R_1(u)$ and $R_2(u)$ are polynomials in $u$ of degree $g$,
with highest degree monomial $u^g$. The polynomials $R_1(u)$ and $R_2(u)$
may be evaluated 
using an expansion for large $u$. By inspection, the  Abelian differentials 
$R_{1,2}(u)du/s(u)$ are meromorphic with a double pole at $\infty$, and $2g$
zeros. Identifying residues at $u=e_i$ gives,
\bea
A_i & = &   - P(e_i) Q_1(e_i) \, (E_i)^{-2}
\no \\
B_i & = &   - P(e_i) Q_2(e_i) \, (E_i)^{-2}
\eea
Expressing the double poles in terms of differentials of simple poles, using 
(\ref{doublepoles}), gives
\bea
\p h_1 & = & 
- i p_1(u)  {du \over s(u)} -i d q _1 (u)
\no \\
\p h_2 & = & 
- p_2(u)  {du \over s(u)} - d q _2 (u)
\eea
where 
\bea
\label{q1q2}
q _1 (u) =  2 \sum _{i=1} ^{2g+1}  { s(u) A_i \over u-e_i} 
& \hskip 0.7in &
p_1(u) = R_1(u) + \sum _{i=1} ^{2g+1} G_i  (u) A_i
\no \\
q _2 (u) =  2 \sum _{i=1} ^{2g+1}  { s(u) B_i  \over u-e_i} 
&  &
p_2(u) = R_2(u) + \sum _{i=1} ^{2g+1} G_i  (u) B_i
\eea
Here, $p_1$ and $p_2$ are real polynomials of $u$ of degree $2g-1$, and 
$q_1$ and $q_2$ are real algebraic functions of $u$.
As a result, we have explicit formulas for the functions $h_1$ and $h_2$
themselves, 
\bea
\label{hes}
h_1(u) & = & h_1 ^{(0)} +
2  \Im \bigg ( q_1 (u)  \bigg ) + 2 \Im \left ( \int ^u_{u_1}  {p_1(u) du \over s(u)} \right )
\no \\
h_2(u) & = & h_2 ^{(0)}
- 2 \Re \bigg ( q_2 (u)  \bigg ) - 2 \Re \left ( \int ^u _{u_2}  {p_2(u) du \over s(u)} \right )
\eea
where $h_1 ^{(0)}$ and $h_2 ^{(0)}$ are real integration constants and 
$u_1, u_2$ are the origins of integration on the real axis. The remaining Abelian integrals 
are smooth functions everywhere, including at the branch points $e_i$, 
except possibly at the branch point at  $\pm \infty$.

\subsection{Positivity of $h_1$, $h_2$ near the branch points}

Condition (R6) of \ref{sectionreg} requires that  $h_1>0$ and $h_2>0$ 
for all  $\Im (u)<0$. We begin by ensuring their positivity at the branch points $e_i$, 
where the  behavior of the simple pole dominates. Using (\ref{hes}) near
a branch point $u \sim e_i$, we have 
\bea 
h_1 (u) & = & + 2 A_i \, \Im \left (  { s(u) \over u- e_i} \right ) + \cO (1) 
\no \\
h_2 (u) & = & - 2 B_i \, \Re \left ( { s(u) \over u- e_i} \right ) + \cO (1) 
\eea
Some care is needed in analyzing this condition. Starting from $u=+\infty$,
move $u$ along the real axis to the left. On the branch $]e_1, +\infty[$, we define 
the function $s(u)$ to be negative. (Making the positive choice would flip the signs 
of both $h_1$ and $h_2$, which would be an immaterial change in our set-up.)
This fixes the phase of $s(u)$ for all $u \in {\bf R}$,  as follows,
\bea
\label{sphase}
s(u)/|s(u)| & = & -1  \hskip 0.5in u \in ~ ]e_1, +\infty[ \, \cup \, \bigcup _{j=1}^{n_1}
]e_{4j+1} , e_{4j}[
\no \\
s(u)/|s(u)| & = & +i  \hskip 0.5in u \in  \bigcup _{j=0}^{n_2}
] e_{4j+2} , e_{4j+1} [
\no \\
s(u)/|s(u)| & = & +1  \hskip 0.5in u \in  \bigcup _{j=0}^{n_3}
] e_{4j+3} , e_{4j+2} [
\no \\
s(u)/|s(u)| & = & -i  \hskip 0.5in u \in  \bigcup _{j=0}^{n_4}
] e_{4j+4} , e_{4j+3} [
\eea
The upper limits of these unions are given by
\bea
\label{nvals}
n_1 & = & [g/2] 
\no \\
n_2 & = & [(2g-1)/4] + n_{-\infty}
\no \\
n_3 & = & [(g-1)/2]
\no \\
n_4 & = & [(2g-3)/2] + 1- n_{-\infty}
\eea
where $n_{-\infty} =1$ when $g$ is even and $n_{-\infty}=0$ when $g$ is odd,
and $e_{2g+2}=-\infty$.

\medskip

We now turn to evaluating the signs of the pole contributions to $h_1$ and $h_2$.
Approaching the branch points $e_{2j}$ from the left and the branch points 
$e_{2j-1}$ from the right yields zero contribution to $h_1(u)$, because the ratio
$s(u)/(u-e_i)$ is real there, while approaching the branch points $e_{2j}$ from the right 
and the branch points $e_{2j-1}$ from the left yields zero contribution to $h_2(u)$, 
because the ratio $s(u)/(u-e_i)$ is imaginary there. 

\medskip

Non-vanishing contributions are obtained for $h_1$ when $e_{2j}$
is approached from the right and $e_{2j-1}$ from the left. For $h_1$, 
approaching $e_{4j+2}$ from the right and $e_{4j+1}$ from the left gives 
$s/|s|= +i$; while approaching  $e_{4j+4}$ from the right and $ e_{4j+3}$ 
from the left  gives  $s/|s|=-i$.  For $h_2$, approaching $e_{4j+2}$
from the left and $e_{4j+3}$ from the right gives $s/|s|= +1$; while approaching 
$e_{4j+4}$ from the left and $ e_{4j+1}$ from the right  gives  $s/|s|=-1$. 
These values suffice to establish the signs required on $A_i$ and $B_i$ to
make $h_1, h_2 >0$ at the branch points, and we find, 
\bea
A_{4j} <0 & \hskip 1in & B_{4j} <0 
\no \\
A_{4j+1} <0 & \hskip 1in & B_{4j+1} > 0
\no \\
A_{4j+2} >0 & \hskip 1in & B_{4j+2} > 0
\no \\
A_{4j+3} >0 & \hskip 1in & B_{4j+3} < 0
\eea
Using the fact that we always have $P(e_i)>0$, it is straightforward to translate 
these conditions into equivalent conditions of the 
polynomials $Q_1$ and $Q_2$, and we find,
\bea
\label{ordere}
Q_1 (e_{4j}) > 0 & \hskip 1in & Q_2(e_{4j}) > 0 
\no \\
Q_1(e_{4j+1}) > 0 & \hskip 1in & Q_2(e_{4j+1})<  0
\no \\
Q_1(e_{4j+2})  < 0 & \hskip 1in & Q_2(e_{4j+2})< 0
\no \\
Q_1(e_{4j+3}) < 0 & \hskip 1in & Q_2(e_{4j+3})> 0
\eea
Together with the ordering of the branch points $e_i$ and the zeros of $Q_1$ and $Q_2$
these conditions give us the possible relative orderings of these points. For higher genus,
there are clearly many combinatorial possibilities.

\smallskip

Of course, having proven that $h_1, h_2 >0$ near the branch points for the above 
assignments does not prove that positivity holds throughout the lower half-plane.
This result requires first solving the regularity condition (R7) of subsection \ref{sectionreg},
which we do in the next subsection.

\subsection{Period relations}
\label{sectionperiodrelations}

Condition (R7) of subsection \ref{sectionreg}, requires that wherever 
$h_1$ or $h_2$ satisfies Dirichlet conditions, it actually must vanish there: 
all Dirichlet boundary conditions must actually be {\sl vanishing}. 
Thus, we must require that $h_1$ and $h_2$ satisfy vanishing Dirichlet boundary 
conditions on $\p \Sigma _+$ and  $\p \Sigma _-$ respectively. 

\medskip

We begin by requiring  $h_1=0$ on the Dirichlet segment 
$[e_1, +\infty] \subset \p \Sigma _+$, and $h_2=0$ on the Dirichlet segment 
$[- \infty, e_{2g+1}] \subset \p \Sigma _-$.
As a  result, the integration constants in (\ref{hes}) get fixed, and we have\footnote{To 
properly define these integrals up to infinity,  one should first introduce a cutoff,
and then take the limit. No contributions to $h_1$ or $h_2$ arise from these 
segments, however, whether finite or infinite, because we take the real
or imaginary parts.}
\bea
\label{h1h2}
h_1(u) & = & 
2  \, \Im \bigg ( q_1 (u)  \bigg ) + 2 \, \Im \left ( \int ^u_{+\infty}  {p_1(u) du \over s(u)} \right )
\no \\
h_2(u) & = & 
- 2 \, \Re \bigg ( q_2 (u)  \bigg ) - 2 \, \Re \left ( \int ^u _{-\infty }  {p_2(u) du \over s(u)} \right )
\eea
Here, we have also used the fact that $q_1$ is manifestly  real on $[e_1, +\infty] $,
and $q_2$ manifestly imaginary on $[- \infty, e_{2g+1}]$.
To guarantee that $h_1(u)=0$ also when $u \in [e_{2j+1},e_{2j}]$, for $j=1,\cdots,g$, 
and $h_2(u)=0$ for $u \in [e_{2j}, e_{2j-1}]$ for $j=1,\cdots , g$,
it will suffice  to have 
\bea
\label{h1h2B}
h_1( e_{2j})-h_1(e_{2j-1}) =0 & \hskip 0.5in & j=1,\cdots, g
\no \\
h_2( e_{2j+1})-h_2(e_{2j}) =0 & \hskip 0.5in & j=1,\cdots, g
\eea
The contributions from $q_1$ and $q_2$ vanish provided the branch points 
are approached from a suitable direction (the path of integration of
the differentials $\p h_1$ and $\p h_2$ has to be smooth). The remaining
conditions (\ref{h1h2B}) amount to the following period relations,
\bea
\label{periodh1h2}
\Im \left ( \int ^{e_{2j-1}} _{e_{2j}}  {p_1(u) du \over s(u)} \right ) =0
& \hskip 0.6in & j= 1, \cdots , g
\no \\
\Re \left ( \int ^{e_{2j}} _{e_{2j+1}}  {p_2(u) du \over s(u)} \right ) =0
& \hskip 0.6in & j= 1, \cdots , g
\eea
giving a system of $2g$ real linear relations on the coefficients of 
the polynomials $p_1$ and $p_2$.

\subsection{Positivity of $h_1$ and $h_2$ throughout $\Sigma$}

Combining the results of positivity near the branch points, obtained in (\ref{ordere}),
with the vanishing of $h_1$ and $h_2$ on their respective Dirichlet segments,
we can now derive a further positivity condition for $h_1$ and $h_2$.
It stems from the fact that if $h_1$ and $h_2$ vanish on their Dirichlet segments 
and are to be positive in the upper half-plane, then their normal derivatives
on the real axis must be negative, namely
\bea 
\Im \left ( \p _u h_1 \right ) \bigg | _{\Im (u) =0} < 0 & \hskip 1in & u \in \p \Sigma _+
\no \\
\Im \left ( \p _u h_2 \right ) \bigg | _{\Im (u) =0} <0 && u \in \p \Sigma _-
\eea
Using the fact that $P(u)>0$ for all $\Im (u)=0$, and the phase values of $s$
along the real axis, given in (\ref{sphase}), we obtain the following inequalities
for $Q_1$ and $Q_2$,
\bea
\label{Qsign}
Q_1(u) >0  & \hskip 0.5in &  u \in [e_1, +\infty] \, \cup \, \bigcup _{j=1}^{n_1}
[e_{4j+1} , e_{4j}]
\no \\
Q_2(u) <0  & \hskip 0.5in & u \in  \bigcup _{j=0}^{n_2}
[e_{4j+2} , e_{4j+1}]
\no \\
Q_1(u) <0  & \hskip 0.5in & u \in  \bigcup _{j=0}^{n_3}
[e_{4j+3} , e_{4j+2}]
\no \\
Q_2(u) >0  & \hskip 0.5in & u \in  \bigcup _{j=0}^{n_4}
[e_{4j+4} , e_{4j+3}]
\eea
with the same assignments for $n_1, \cdots, n_4$ as in (\ref{nvals}). These 
conditions imply the inequalities at the branch points in (\ref{ordere}),
but they are actually stronger, as they imply that neither $Q_1$ nor $Q_2$
can have any zeros in their respective Dirichlet segments.

\subsection{The general hyperelliptic solution}

First, we shall show that the sign conditions on $Q_1(e_i)$ and $Q_2(e_i)$ 
put a lower bound on the number of real zeros. The existence of a lower bound 
follows  from the fact, in (\ref{Qsign}),  that the sign of $Q_1(e_k)$ 
and $Q_2(e_k)$ alternates in $k$ with periodicity  4. This alternation requires the 
degrees of the polynomials to satisfy $q={\rm deg} (Q_1)= {\rm deg}(Q_2) \geq g$.
Using the relation $3g+1=2p+q$ of (\ref{PQ1Q2}), 
where $p$ is the number of complex zeros in the lower half-plane, 
it is clear that the  bound $q=g$ can never
be attained, and we have a more stringent bound,
\bea
\label{lowerq}
g +1 \leq q={\rm deg} (Q_1)= {\rm deg}(Q_2) 
\eea
which in turn implies that the number $p$ of complex zeros $u_a$ obeys $p \leq g$. 

\smallskip

Second, we shall use the fact, derived in (\ref{Qsign}),  
that $Q_1$ and $Q_2$ have no zeros on their respective
Dirichlet segments, so that, 
\bea
\label{rootsets}
\{ \a_1, \cdots , \a _q \} & \subset & \cU_-
\no \\
\{ \b _1 , \cdots , \b_q \} & \subset & \cU_+
\eea
as well as the fact that the condition $W<0$ requires  the zeros
of $Q_1$ and $Q_2$ to alternate as in (\ref{alter}). 
Since each elementary interval $[e_1, +\infty]$ and $[e_{i+1}, e_i]$ for $i=1,\cdots, 2g-1$ 
either contains no $\alpha$ roots or no $\beta $ roots, we conclude that it cannot
contain more than one root (whether $\a$ or $\b$). Suppose, for example,  that the 
interval $[e_{4j+2}, e_{4j+1}]$ contained two consecutive $\alpha$ roots as (\ref{alter}), 
then it would also have to contain the $\beta$ root which lies in between the two
$\alpha $ roots and this is not allowed by (\ref{Qsign}). Thus, each elementary 
interval contains at most one root. This gives an upper bound on the total 
number of real roots, 
\bea
\label{upperq}
2 q \leq 2g +2
\eea
which we simply obtain from counting the total number of elementary intervals,
$2g+2$.  Combining the lower bound on $q$ in (\ref{lowerq})
and the upper bound on $q$ in (\ref{upperq}), it is immediate that we must have
\bea
p & = & g 
\no \\
q & = & g+1 
\eea
which implies a {\sl unique relative ordering of the roots and branch points}, 
given by
\bea
\label{totalorder}
\alpha _{g+1} < e_{2g+1} < \beta _{g+1} < e_{2g}  < \cdots 
< \alpha_b < e_{2b-1} < \beta _b < 
\cdots  < e_2 < \alpha _1 < e_1 < \beta _1 
\quad
\eea
for $g \geq b \geq 2$. Once this ordering is satisfied, conditions (R5) and (R6) 
will be obeyed.

\subsection{Summary of the hyperelliptic solution}

In summary, the genus $g$ hyperelliptic Riemann surface $\Sigma$ is represented by
the lower half-plane, and its boundary $\p \Sigma$ is represented by the real line.
The $2g+2$ branch points $e_1 , \cdots , e_{2g+1}, e_{2g+2}=\infty$ are real.
The harmonic functions are defined by their differentials, $\p h_1$ and $\p h_2$,
given by 
\bea
\label{diffh1h2a}
\p  h_1  & = & - i \, { P(u) Q_1(u)\over s(u)^3} du
\no \\
\p  h_2  & = &  - { P(u) Q_2(u) \over s(u)^3} du
\eea
Here, $s$ was given in (\ref{seq}); $P$ is a polynomial of degree $2g$ in $u$, 
whose zeros come in complex pairs $(u_a, \bar u_a)$, with $\Im (u_a)<0$; 
and $Q_1$ and $Q_2$ are polynomials of degree $g+1$ in $u$ with real roots 
$\a_1 , \cdots, \a_{g+1}$ and  $\b_1 , \cdots , \b_{g+1}$ respectively. 
The form of the differentials $\p h_1$ and $\p h_2$, and the relative ordering 
(\ref{totalorder})  between the real zeros and the branch points guarantees 
that conditions (R1-R7) are satisfied provided we impose also the period
relations 
\bea
\Im \left ( \int _{e_{2j}} ^{e_{2j-1}} \p h_1 \right ) = 0
& \hskip 1in & j=1, \cdots , g
\no \\
\Im \left ( \int _{e_{2j+1}} ^{e_{2j}} \p h_2 \right ) = 0 &&  j=1, \cdots , g
\eea
Given a set of real zeros and branch points with  ordering (\ref{totalorder}), 
the $2g$ period relations determine the $g$ complex zeros $u_a$.

\smallskip

Solutions of these period relations may not exist for given moduli of $\Sigma$, 
and all values of the real roots $\alpha$ and $\beta$ consistent with (\ref{totalorder}). 
In the section 5, the allowed parameter space will be explored analytically for 
the genus 1 case, and it will be shown that non-singular solutions do indeed 
exist for open sets of the full parameter space. In section 6, the existence 
of regular solutions will be explored also for higher genus, in part by analytical
and in part by numerical methods, and it will be shown for genus 2 that regular 
solutions exist in an open set of the full parameter space.

\subsection{Asymptotic behavior near the branch points}\label{asymbranch}

In this subsection, the asymptotic behavior of the above non-singular solutions 
near a branch point $u=e_{i}$ will be analyzed.  Defining $u= e_{i} + z$, the 
harmonic functions $h_{1},h_{2}$ behave as follows near $z=0$,
\bea
h_{1}&=& 
2i \left ( \gamma_{1} {1\over \sqrt{z}  }- \delta_{1}  \sqrt{z} \right ) 
+ o \left ( z^{3\over 2} \right )
+{\rm c.c.} \no \\
h_{2}&=& 
2 \left ( \gamma_{2}{1  \over  \sqrt{z} }- \delta_{2} \sqrt{z} \right )  
+ o \left ( z^{3\over 2} \right )
+{\rm c.c.} 
\eea
where the constants are easily obtained from (\ref{diffh1h2})
\bea
\gamma_{1} &=& {P(e_{i}) Q_{1}(e_{i}) \over \prod_{j\neq i} (e_{i}-e_{j})^{3/2}}
\no\\ 
 \gamma_{2} &=& {P(e_{i}) Q_{2}(e_{i}) \over \prod_{j\neq i} (e_{i}-e_{j})^{3/2}} 
\eea
and 
\bea
\delta_{1}&=& \gamma_{1} \left ( 
\sum_{k}^{p} \big({1\over e_{i } -u_{k}} +{1\over e_{i } -\bar u_{k}}\big) 
+\sum_{k}^{q} {1\over q_{i }-\alpha_{k}} 
-{1\over 3} \sum_{k\neq i}^{2g-1}{1\over e_{k}-e_{i}} \right ) 
\no \\
\delta_{2}&=& \gamma_{2} \left ( 
\sum_{k}^{p} \big({1\over e_{i } -u_{k}} +{1\over e_{i } -\bar u_{k}}\big) 
+\sum_{k}^{q} {1\over q_{i }-\beta_{k}} 
-{1\over 3} \sum_{k\neq i}^{2g-1}{1\over e_{k}-e_{i}} \right )
\eea
Comparison with the genus 0 Janus solution (\ref{janushalf}) reveals 
that near each of the branch points the supergravity geometry approaches 
$AdS_{5}\times S^5$ asymptotically.  Introducing a new coordinate 
$z= e^{-2 x- 2 i y}$ the asymptotic  region near the branch point  $z=0$ 
is mapped to $x\to \infty$ and the  coordinate $y\in[0,\pi/2]$.

\subsubsection{The dilaton and the metric}

The asymptotic values of the dilaton near each  $i$-th branch point $e_{i}$ 
is given by
\bea\label{asymdil}
e^{2 \phi} = \left | {Q_2(e_i) \over Q_1(e_i)}\right| + o(e^{-4 x})
\eea
which are constants which depend on $e_i$, $\a_b$ and $\b_b$. 
The metric becomes
\bea\label{asymmet}
ds^{2}= 4 \sqrt{2 \Delta }\Big( dx^{2 }+dy^{2} + (\cos y)^2 ds_{S_{1}}^{2} 
+ (\sin y)^2 ds_{S_{2}}^{2}  +   e^{2x} ds_{AdS_{4}}^{2}\Big)+o(e^{-2x})
\eea
where 
\be
\Delta= \delta_{2}\gamma_{1}-\delta_{1}\gamma_{2}
= \gamma_{1}\gamma_{2} 
\left (\sum_{k=1}^{q} {1\over e_{i} -\beta_{k}} -{1\over e_{i} -\alpha_{k}} \right )
\ee
In the limit $x \to \infty$ the metric becomes  $AdS_{5} \times S^{5}$, and 
$4 \sqrt{2\Delta}$ is the radius squared  of the $AdS_5 \times S^5$ 
geometry in the neighborhood of $e_i$.
Since there are  $2g+2$ branch points (including infinity), the genus $g$ 
solution has $2g+2$ asymptotically $AdS_{5}\times S^5$ regions, where 
the dilaton approaches (generically) different constant values. 
The holographic interpretation of this 
geometry will be presented in section \ref{holdual}.

\subsubsection{The 2-form potential $B_{(2)}$}

The functions $b_{1},b_{2}$ (\ref{bsol2}) parametrize 
the NSNS and RR  2-form potentials of $B_{(2)}$. To compute $b_1$ and $b_2$,  
we need to evaluate 
the harmonic duals $\tilde h_1$ and $\tilde h_2$. Recall 
equation (9.24) of \cite{degAdS4} for the harmonic functions $h_1$ and $h_2$,
and equation (9.43) for their harmonic duals $\tilde h_1$ and $\tilde h_2$,
in terms of the holomorphic functions $\cA$ and $\cB$, 
\bea
h_1 = -i (\cA - \bar \cA )  & \hskip 1in & \tilde h_1 = \cA + \bar \cA
\no \\
h_2 = \cB + \bar \cB \hskip 0.32in && \tilde h_2 = i (\cB - \bar \cB )
\eea
The holomorphic functions are found as follows,
\bea
\cA & = & \cA_0 - 2 \Big( \gamma_{1} {1\over \sqrt{z}  }- \delta_{1}  \sqrt{z} \Big)
+ o ( z^{3 \over 2} )
\no \\
\cB & = & 
i \cB_0 + 2 \Big( \gamma_{2}{1  \over  \sqrt{z} }- \delta_{2} \sqrt{z} \Big)  
+ o(z^{3\over 2})
\eea 
Here, $\cA_0$ and $\cB_0$ are real $z$-independent parameters.
They arise because the splitting of the harmonic functions $h_1$ and $h_2$ into
holomorphic ones is unique only up to additive constants, which cannot
be determined from the local properties of $h_1$ and $h_2$.
(In the next subsection, the difference between the values of these constants
at different branch points will be determined using Abelian integrals.)
Putting all together, we have now the following asymptotic expressions 
for the fields $b_1$ and $b_2$,
\bea
b_{1}&=& - 4 \cB_0 
+ {32 \Delta (\delta_{2 }\gamma_{1} + \delta_{1}\gamma_{2} )\over \gamma_{1}^{2} \gamma_{2}}\; (\sin y)^{3}   e^{-3x}+ o(e^{-5x})
\no \\
b_{2}&=& - 4 \cA_0 
+{32 \Delta (\delta_{2 }\gamma_{1} + \delta_{1}\gamma_{2}) \over \gamma_{2}^{2} \gamma_{1}}\; (\cos y)^{3}   e^{-3x}+ o(e^{-5x})
\eea
The associated 3-form fluxes vanish in the asymptotic $AdS_5 \times S^5$ regions.

\subsection{Homology 3-spheres}

The hyperelliptic solutions exhibit non-trivial 3-cycles on which the 3-form fields 
have non-zero charges. In this section, we determine
these 3-cycles and evaluate the 3-form charges. 

\smallskip

A non-trivial 3-cycle arises when a 1-parameter family of 2-spheres, 
either $S_1^2$ or $S_2^2$, starts at zero radius (respectively $f_1$ or $f_2$)
and returns to zero radius in a manner consistent with the topology
of 3-spheres (respectively $S_1^3$ or $S_2^3$).  The relevant 1-parameter 
families correspond to intervals on the real line $\p \Sigma$, located between 
consecutive branch points. The precise correspondence is as follows:
\bea
S_{1j} ^3  & = & \left \{ [e_{2j}, e_{2j-1}] \times_f S_1^2 \right \} 
\hskip 1in j=1, \cdots, g
\no \\
S_{2j} ^3 & = & \left \{ [e_{2j+1}, e_{2j}] \times_f S_2^2 \right \}
\hskip 1in j=1, \cdots, g
\eea
Here, it is understood that the product $\times_f$ of the branch cut and a 
2-sphere stands for a fibration of the 2-sphere over the interval, and not 
for a product of sets.
Notice that for genus~0, which corresponds to the $AdS_5 \times S^5$ and Janus 
solutions, no non-trivial 3-cycles are found to exist.

\subsection{Evaluation of the RR and NSNS 3-form charges}

The real NSNS and RR 3-form field strengths $H_{(3)}$ and $C_{(3)}$,
are differentials of the complex 2-form potential, 
given by the relation $H_{(3)} + i C_{(3)} = d B_{(2)}$.
The explicit expression for $B_{(2)}$ on our solutions is given by 
(\ref{bsol1}) and (\ref{bsol2}). As a result, the charges $\cH_j$ and $\cC_j$, respectively of the fields $H_{(3)}$ and $C_{(3)}$,  across the non-trivial 3-cycles 
$S_{1j}^3$ and $S_{2j}^3$ are given as follows,
\bea
\label{HC}
\cH _j & \equiv & 
\int _{S_{1j}^3} db_1 \wedge \hat e^{45} = + 8 \pi \int _{e_{2j}} ^{e_{2j-1}} d \tilde h_2
\no \\
\cC _j & \equiv & 
\int _{S_{2j}^3} db_2 \wedge \hat e^{67} = - 8 \pi \int _{e_{2j+1}} ^{e_{2j}} d \tilde h_1
\eea
In using the relation (\ref{bsol2}) between $b_{1,2}$ and $\tilde h _{1,2}$
in the above expressions, the second terms on the right hand sides of (\ref{bsol2})
cancel because $h_1=0$ on the intervals entering into the calculation of $\cH_j$,
while $h_2=0$ on the intervals entering the calculation of $\cC_j$. 
An alternative way to see that the second terms on the right hand sides of (\ref{bsol2}) do not contribute is by observing that the above line
integrals are effectively around closed curved on the full hyperelliptic
Riemann surface (including the upper half plane and both Riemann sheets),
and that those contributions to $b_1$ and $b_2$ are single-valued,
and thus cancel out of the integrals.

\smallskip

To evaluate the line 
integrals over $d \tilde h_{1,2}$, we use (\ref{tildes}) to express the differentials
of $\tilde h_{1,2}$ in terms of those of $h_{1,2}$, and then use (\ref{h1h2}) 
to evaluate the line integrals. The functions $q_1(u)$ and $q_2(u)$ of 
(\ref{q1q2}), which enter in the expression (\ref{h1h2}), are single-valued
scalars and do not contribute to the line integrals of (\ref{HC}). The remaining
integrals give,
\bea
\cH _j & = & - 16 i \pi \int _{e_{2j}} ^{e_{2j-1}} {p_2(u) \, du \over s(u)}
\no \\
\cC _j & = & - 16 \pi  \int _{e_{2j+1}} ^{e_{2j}} { p_1(u) \, du \over s(u)}
\eea
Both integrals are real, given the phase of $s(u)$ in (\ref{sphase}), and 
generically non-vanishing.

\newpage

\section{Genus 1 solutions}
\setcounter{equation}{0}

The elliptic case provides the simplest solution of the 
hyperelliptic Ansatz that goes beyond the $AdS_5 \times S^5$ and Janus
solutions. Generically, it will have four distinct asymptotic $AdS_5 \times S^5$
regions, each with a different value of the dilaton. The Abelian integrals may be recast
in terms of the familiar elliptic functions on the torus,\footnote{Useful general 
references on elliptic functions, conformal mapping, and explicit formulas may be 
found in the Bateman manuscript \cite{bateman}, and in \cite{magnus}.}
and the domain of  parameter space that leads to non-singular solutions may be 
constructed  explicitly and analytically.

\subsection{Formulation on the lower half-plane}

First, the parametrization given in the preceding section for all genera 
simplifies considerably at genus 1, and reduces to the following,
\bea
s(u)^2 & = & (u-e_1)(u-e_2)(u-e_3)
\no \\
P(u) & = & (u - u_1) (u - \bar u_1)
\no \\
Q_1(u) & = & (u -\a_1) (u- \a_2)
\no \\
Q_2 (u) & = & (u - \b_1) (u-\b_2)
\eea
where we may choose $e_1+e_2+e_3=0$, without loss of generality. 
The branch points $e_1, e_2, e_3$  and the roots $\a_1, \a_2,\b_1, \b_2$ 
are real, and subject to the ordering relation (\ref{totalorder}) for $g=1$,
\bea
\label{ordergenus1}
\a_2 < e_3 < \b_2 < e_2 < \a_1 < e_1 < \b_1
\eea
and $\Im (u_1) <0$.
We use the calculations of subsection \ref{sectionh1h2} to derive the 
period relations in the elliptic case. To this end, we compute
$G_i(u) = -u + e_i$, as well as the following objects,
\bea
\label{p1p2}
p_1(u) & = & u- A_4 - \sum _{i=1}^3 A_i (u-e_i)
\no \\
p_2(u) & = & u - B_4 - \sum _{i=1}^3 B_i (u-e_i)
\eea
The constants $A_i, B_i$  are obtained as residues and are given by
\bea
A_i & = &   - P(e_i) Q_1(e_i) \, (E_i)^{-2}
\no \\
B_i & = &   - P(e_i) Q_2(e_i) \, (E_i)^{-2}
\eea
while the constants $A_4$ and $B_4$ may be obtained from the next-to-leading order
behavior at $u = \infty$ and are found to be, 
\bea
A_4 & = & \a_1 + \a_2 + u_1 + \bar u_1
\no \\
B_4 & = & \b_1 + \b_2 + u_1 + \bar u_1
\eea
Here, $E_i$ is given solely in terms of the branch points by $E_i \equiv 
(e_i - e_j)(e_i-e_k)$ where $e_j$ and $e_k$ are two distinct branch points
which are distinct also from $e_i$.
The period relations of (\ref{periodh1h2}) take the following form, 
\bea
\Im \left ( \int ^{e_1} _{e_2}  {p_1(u) du \over s(u)} \right ) & = & 0
\no \\
\Re \left ( \int ^{e_2} _{e_3}  {p_2(u) du \over s(u)} \right ) & = & 0
\eea
They may be rendered more  explicit by using the expressions for $p_1$ and $p_2$
of (\ref{p1p2}), as  well as the following basic elliptic integrals, 
\bea
\int  ^{e_1} _{e_2}  { du \over s(u)} = \omega _3
&\hskip 1in &
\int  ^{e_1} _{e_2}  { u \, du \over s(u)} = - \zeta (\omega _3)
\no \\
\int  ^{e_2} _{e_3}  { du \over s(u)} = \omega _1
&\hskip 1in &
\int  ^{e_2} _{e_3}  { u\, du \over s(u)} = - \zeta (\omega _1)
\eea
where $\zeta (u)$ is the Weierstrass $\zeta$-function. In view of (\ref{sphase}),
the periods $\omega _1$ and $\zeta (\omega _1)$ are real, while 
the periods $\omega _3$ and $\zeta (\omega _3)$ are purely imaginary.
The  period relations may then be recast as follows,
\bea
\label{periods1}
A_4 =  - \zeta  _3 + \sum _{i=1} ^3 A_i  \left (e_i + \zeta  _3 \right )
& \hskip 1in & 
\zeta _3 \equiv {\zeta (\omega _3) \over \omega _3} 
\no \\
B_4 =  - \zeta  _1 + \sum _{i=1} ^3 B_i  \left ( e_i + \zeta  _1 \right )
& \hskip 1in & 
\zeta _1 \equiv {\zeta (\omega _1) \over \omega _1} 
\eea
Together with the defining relations for $A_i, B_i, A_4$ and $B_4$, 
the period relations give two real equations which may be viewed as
equations for the complex zero $u_1$, as a function of the real modulus 
of the torus (parametrized by one of the branch points $e_1, e_2, e_3$)
and the real zeros $\a_1, \a_2, \b_1, \b_2$. The relations  implied
on $u_1$ are quadratic and of the following form, 
\bea
\label{circles}
a_0 |u_1|^2 - a_1 (u_1 + \bar u_1) + a_2 &=& 0 \hskip 1in \Im(u_1)<0
\no \\
b_0 |u_1|^2 - b_1 (u_1 + \bar u_1) + b_2 &=& 0
\eea
where the coefficients of these quadrics are given by
\bea
\label{ab}
a_n & = & - \delta _{n,1} + (\a_1+\a_2+\zeta _3) \delta _{n,2}
+ \sum _{i=1}^3 e_i ^n (e_i + \zeta _3) Q_1(e_i) E_i ^{-2}
\no \\
b_n & = & - \delta _{n,1} + (\b_1+\b_2+\zeta _1) \delta _{n,2}
+ \sum _{i=1}^3 e_i ^n (e_i + \zeta _1) Q_2(e_i) E_i ^{-2}
\eea
for $n=0,1,2$ and $\delta _{n,1}$ and $\delta _{n,2}$ are the Kronecker $\delta$.
The quadrics of (\ref{circles}) are two half-circles whose centers lie on the real axis.

\subsection{Parameter space via four real zeros}

To investigate the existence of solutions to (\ref{circles}) and (\ref{ab}), 
it is convenient
to recast (\ref{circles}) in terms of the centers and radii of the circles,
\bea
\label{circles1}
\left | u_1 - {a_1 \over a_0} \right |^2 = R_a^2
& \hskip 1in & R_a ^2 = {a_1^2 - a_0 a_2 \over a_0^2}
\no \\
\left | u_1 - {b_1 \over b_0} \right |^2 = R_b^2
& \hskip 1in & R_b ^2 = {b_1^2 - b_0 b_2 \over b_0^2}
\eea
The necessary and sufficient conditions for the existence of a solution
are that these circles have positive $R_a^2$ and $R_b^2$, and have 
non-trivial intersection. 
When they do, the intersection will produce a single point $u_1$, which is the 
unique common complex zero of the differentials $\p h_1$ and $\p h_2$
in the lower half-plane. In summary, these conditions are 
\bea
\label{circles2}
0  \leq  R_a^2 ,  R_b^2 
\hskip 1in \left | { a_1 \over a_0}  - {b_1 \over b_0} \right |  \leq  R_a + R_b
\eea
In the next subsection, we shall show that the coefficients $a_n, b_n$ may be expressed in terms of genus 1 modular forms.

\subsection{Parameter space via the complex and two real zeros}

A more practical approach is to use the complex zero itself as a parameter,
together with the two real zeros $\a_1, \b_2$ whose range is compact.
We may then solve for $\a_2$ and $\b_1$, using the relations (\ref{circles})
together with (\ref{ab}). One obtains,
\bea
\a_2 & = & 
{\a_1 + u_1 + \bar u_1 + \zeta _3 + \sum _{i=1}^3 (e_i + \zeta _3) e_i (e_i - \a_1) 
|u_1-e_i|^2 E_i ^{-2}
\over 
-1 + \sum _{i=1}^3 (e_i + \zeta _3)  (e_i - \a_1)  |u_1-e_i|^2 E_i ^{-2}}
\no \\
\b_1 & = & 
{\b_2 + u_1 + \bar u_1 + \zeta _1 + \sum _{i=1}^3 (e_i + \zeta _1) e_i (e_i - \b_2) 
|u_1-e_i|^2 E_i ^{-2}
\over 
-1 + \sum _{i=1}^3 (e_i + \zeta _1)  (e_i - \b_2)  |u_1-e_i|^2 E_i ^{-2}}
\eea
subject to the range (\ref{ordergenus1}). Considering the modulus as given, then
for each set of values $\a_1$ and $\b_2$ in the range (\ref{ordergenus1}),
the above expression, together with the ranges for $\a_2$ and $\b_1$, 
limit the domain of $u_1$ in the complex plane by two semi-circles. 
The allowed range of the parameters will be obtained analytically
for the case of the square torus, where many simplifications occur.

\begin{figure}[tbph]
\begin{center}
\epsfxsize=3.0in
\epsfysize=3.0in
\epsffile{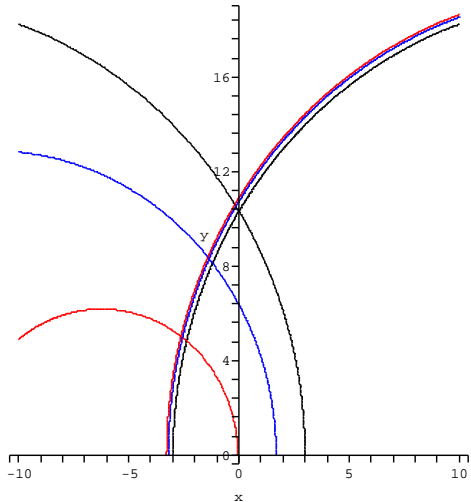}
\label{figure10}
\caption{Allowed parameter region for the complex zero $u_1=x+iy$
for given values of $\a_1$ and $\b_2$ (equal to $(e_1+e_2)/2$ and 
$(e_2+e_3)/2$ in this case), and varying $\tau = i$ (black arcs), $1.2 i$ 
(blue arcs),  and $1.6i$ (red arcs). The allowed regions are the trigons 
bounded by the real axis  and arcs of  the same color.}
\end{center}
\end{figure}

\subsection{Formulation in terms of elliptic functions and explicit solution}

An entire genus 1 Riemann surface, i.e. the double cover of the plane, is 
uniformized by the Weierstrass  function which maps  the torus\footnote{We adopt the conventions of  \cite{bateman} with $\omega = \omega _1$ real, and 
$\omega ' = \omega _3$, and $\tau = \omega _3 /\omega _1$ purely imaginary.} 
with half-periods $\omega _1$ and $\omega _3$,
and modulus  $\tau = \omega _3 /\omega _1$ into the double cover of the plane 
by the map
\bea
z \longrightarrow (\wp (z), \wp'(z))
\eea
where $\wp'(z)$ is given in terms of $\wp(z)$ through the defining equation,
\bea
\label{defwp}
\wp'(z)^2 = 4 (\wp (z) - e_1) (\wp (z) - e_2) (\wp(z) - e_3)
\eea
up to a sign. This sign distinguishes between the upper and lower Riemann 
sheets. The branch points are related to the half-periods by 
$e_i = \wp (\omega _i)$, for $i=1,2,3$,  and $\infty = \wp (0)$.

\begin{figure}[tbph]
\begin{center}
\epsfxsize=6in
\epsfysize=2.5in
\epsffile{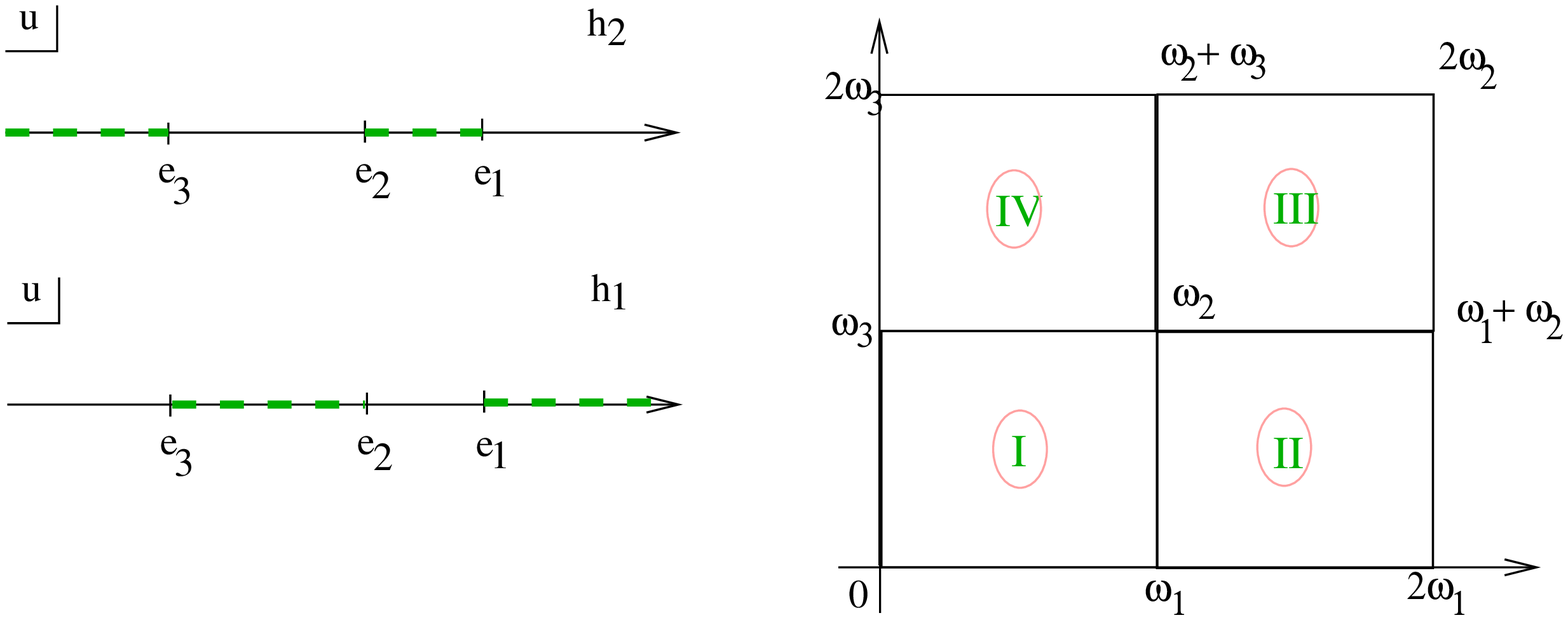}
\label{figure3}
\caption{The cut plane and fundamental domain of the elliptic solution.}
\end{center}
\end{figure}

The Weierstrass function  $\wp (z)$ is periodic with periods $2 \omega _1$ 
and  $2 \omega _3$. It is real if and only if either $z= m \omega _1 + i \rho$, 
or $ z= n \omega _3 + \rho$, both  with $ \rho$ real, and $m$ and $n$ arbitrary integers.
This divides the fundamental region for the torus into 4 regions (see Figure 4). 
Under $z \to 2 \omega _1 + 2 \omega _3 -z$, the  torus is mapped into itself,  
$\wp \to \wp$
but $\wp ' \to - \wp'$, so that the two Riemann sheets are interchanged,
and regions I and III, as well as II and IV, are interchanged with one another..
Furthermore, regions I and III map to the lower half-plane, while regions II and IV
map to the upper half-plane. The lower half-plane of a single sheet is the image of region I. 
Using the map $u = \wp(z)$, the inversion formulas \cite{bateman},
\bea
\wp(z+\omega _i) = e_i + {E_i \over \wp(z) - e_i} \hskip 1in i=1,2,3
\eea
and the sign choice $s(u)= \wp'(z)/2$, the differentials $\p h_1$ and $\p h_2$
may be recast in terms of the Weierstrass function,
\bea
\label{diffs}
\p h_1 & = & 
i \sum _{\alpha =0} ^3 
A_\alpha \bigg ( \wp (z+\omega _\alpha) +{\zeta  _3} \bigg )dz
\no \\
\p h_2 & = & 
 \sum _{\beta =0} ^3 B_\beta \bigg ( \wp (z+\omega _\beta) +{\zeta  _1} \bigg )dz
\eea
where $A_0= B_0 =-1$. With the help of the relation between the Weierstrass 
functions, $\zeta '(z) = - \wp (z)$, it is immediate to derive the expressions 
of the harmonic functions $h_1$ and $h_2$ themselves. The overall
additive integration constant generated in the process is fixed 
by the requirement of vanishing Dirichlet boundary conditions, 
on the segments $[0, \omega _1]$ and $[\omega _3, \omega _2]$
for $h_1$ and on the segments $[0, \omega _3]$ and $[\omega _1, \omega _2]$
for $h_2$, and we find,
\bea
\label{ellhes}
h_1(z) & = & 
2 i   \zeta (\omega _3) (A_2+A_3)
- \sum _{\alpha =0}^3 i A_\alpha 
	\left [ \zeta (z + \omega _\alpha) - \zeta (\bar z + \bar \omega _\alpha)
	- (z-\bar z)  {\zeta  _3} \right ]
\no \\
h_2(z) & = & 
2   \zeta (\omega _1) (B_1+B_2)
- \sum _{\alpha =0}^3  B_\alpha 
	\left [ \zeta (z + \omega _\alpha) + \zeta (\bar z + \bar \omega _\alpha)
	- (z + \bar z)  {\zeta  _1} \right ]
\eea
To show the Dirichlet vanishing on the segments $[\omega _3, \omega _2]$
and $[\omega _1, \omega _2]$, we have made use of the following 
addition formula for the $\zeta$-function \cite{bateman}, 
\bea
\label{zetadd}
\zeta (z+z') = \zeta (z) + \zeta (z') + \half {\wp '(z) - \wp '(z')\over \wp (z) - \wp (z')}
\eea
and the fact that $\wp'(\omega _i)=0$ for $i=1,2,3$.

\bigskip 
\bigskip 

\begin{figure}[tbph]
\begin{center}
\epsfxsize=5in
\epsfysize=2.7in
\epsffile{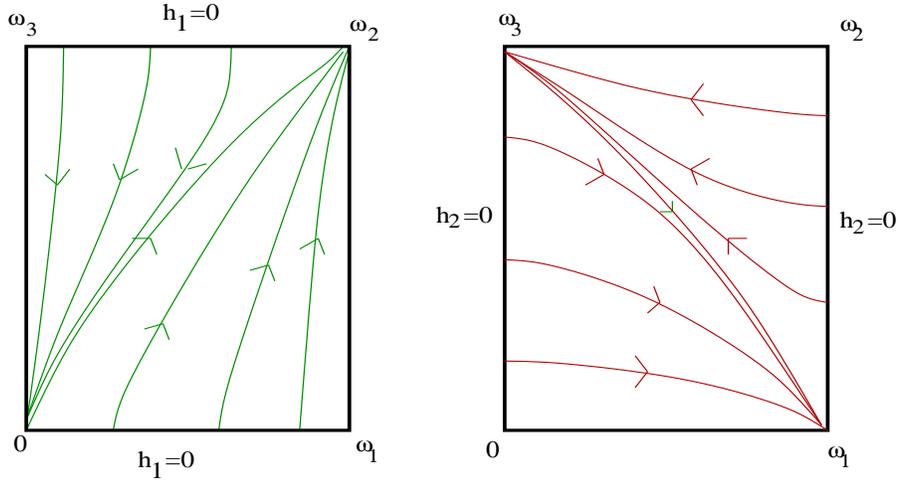}
\label{figure4}
\caption{Field lines on the square of half-periods.}
\end{center}
\end{figure}

\bigskip 

\begin{figure}[tbph]
\begin{center}
\epsfxsize=4.5in
\epsfysize=2in
\epsffile{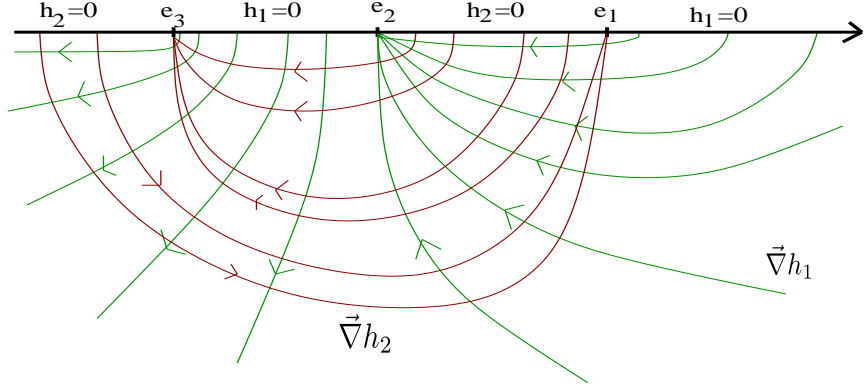}
\label{figure5}
\caption{Field lines on the half-plane.}
\end{center}
\end{figure}

\subsection{Regularity}

The general arguments of the preceding section guarantee that $W<0$
as long as the ordering (\ref{ordergenus1}) holds, and that $h_1$
and $h_2$ are positive on their respective Neumann segments. Positivity 
of $h_1$ and $h_2$ was argued on general grounds as well, but in the 
elliptic case, it may actually be given a solid proof using explicit formulas.
The starting point  is two expansion formulas for the Weierstrass $\zeta$-function,
\bea
\zeta (z) & = & z \zeta  _1 + {\pi \over 2 \omega _1}
\sum _{m=-\infty} ^{+\infty} \cotg \left ( {\pi z\over 2 \omega _1} 
+ m \pi {\omega _3 \over \omega _1}  \right )
\no \\
\zeta (z) & = & z \zeta  _3 + {\pi \over 2 \omega _3}
\sum _{m=-\infty} ^{+\infty} \cotg \left ( {\pi z\over 2 \omega _3} + m \pi {\omega _1 \over \omega _3}  \right )
\eea
The first of these formulas is familiar \cite{magnus}, while the second may be 
obtained from the first by interchanging the half-periods $\omega _1$
and $\omega _3$. The first formula will be applied to $h_2$, the second to $h_1$. 
Remarkably, all dependence on $\zeta (\omega _1)$ and $\zeta (\omega _3)$ 
cancels, and $h_1$ and $h_2$ are expressed as simple  infinite series expansions. 
Using the formula $\cotg u + \cotg \bar u = \sin (u + \bar u)/ |\sin u |^2 $, 
complex conjugate pairs of terms may be combined, to obtain the following formulas,
\bea
\label{sumzeta1}
h_1(z) & = & - {\pi i \over 2 \omega _3} \sum _{\alpha =0} ^3 A_\alpha 
\sum _{m=-\infty} ^{+\infty} 
{ \sin \left ( {\pi \over 2 \omega _3}(z-\bar z + \omega _\alpha 
-\bar \omega _\alpha) \right )
 \over | \sin \left ( {\pi \over 2 \omega _3} (z+\omega _\alpha + 2 m \omega _1) \right ) |^2}
\no \\
h_2(z) & = & - {\pi  \over 2 \omega _1} \sum _{\alpha =0} ^3 B_\alpha 
\sum _{m=-\infty} ^{+\infty} 
{ \sin \left ( {\pi \over 2 \omega _1}(z + \bar z + \omega _\alpha  + \bar \omega _\alpha) \right ) \over | \sin \left ( {\pi \over 2 \omega _1} (z+\omega _\alpha + 2 m \omega _3) \right ) |^2}
\eea
Notice that the $\sin$ factors in the numerators are actually independent of 
the summation index $m$
and may be moved out of the sum over $m$. Since these numerators depend only
on either $z-\bar z$ for $h_1$ and $z+\bar z$ for $h_2$, it is immediate that 
positivity on their Neumann boundary segments actually implies positivity
everywhere in the fundamental region bounded by the half-periods,
which thus proves that $h_1, h_2 >0$ everywhere.

\subsection{Analytical solution and modular polynomials}

The solution (\ref{ellhes}) is completely explicit in terms of the branch points $e_i$,
and the real zeros  $\a_1, \a_2, \b_1,\b_2$,  chosen subject to (\ref{ordergenus1}),
as long as the period relations (\ref{periods1}) are solved, by satisfying (\ref{circles})
with (\ref{ab}). The elliptic parametrization will allow us to render the period
relations more explicit by expressing the parameters $a_n$ and $b_n$ of 
(\ref{ab}) in terms of modular forms. Defining the   modular objects,
\bea
\label{modular12}
M_1 ^{(m)} & = & \sum _{i=1}^3 e_i ^m (e_i + \zeta _3) E_i^{-2}
\no \\ 
M_2 ^{(m)} & = & \sum _{i=1}^3 e_i ^m (e_i + \zeta _1) E_i^{-2}
\eea
the parameters $a_n$ and $b_n$ of (\ref{ab}) may be expressed as follows,
\bea
a_n & = & \a_1 \a_2 M_1 ^{(n)} + (\a_1+\a_2) \left ( \delta _{n,2} - M_1 ^{(n+1)} \right )
-\delta _{n,1} + \zeta _3 \delta _{n,2} + M_1 ^{(n+2)}
\no \\
b_n & = & \b_1 \b_2 M_2 ^{(n)} + (\b_1+\b_2) \left ( \delta _{n,2} - M_2 ^{(n+1)} \right )
-\delta _{n,1} + \zeta _1 \delta _{n,2} + M_2 ^{(n+2)}
\eea
To compute the modular objects $M_1$ and $M_2$, we express the branch
points in terms of genus one $\tet$-functions, using the Thomae formulas \cite{bateman},
\bea
\label{branch1}
	e_1 - e_2 & = & {\pi ^2 \over 4 \omega _1^2} \tet _4^4
	= - {i \pi \over \omega _1^2} \p_\tau \ln {\tet_2 \over \tet_3}
\no \\
	e_1 - e_3 & = & {\pi ^2 \over 4 \omega _1^2} \tet _3^4
	= - {i \pi \over \omega _1^2} \p_\tau \ln {\tet_2 \over \tet_4}
\no \\
	e_2 - e_3 & = & {\pi ^2 \over 4 \omega _1^2} \tet _2^4
	= - {i \pi \over \omega _1^2} \p_\tau \ln {\tet_3 \over \tet_4} \quad 
\eea
The relation $e_1+e_2+e_3=0$ holds in view of the famous Jacobi identity
$\tet _2^4 + \tet _4 ^4= \tet _3^4$. The branch points are modular forms 
under the subgroup of the full modular group $SL(2, \bZ)$ which leaves 
the half-periods invariant. With these results, we readily compute the 
combinations $E_i$,
\bea
E_1^2 (e_2-e_3)^2 = E_2^2 (e_3-e_1)^2 = E_3 ^2 (e_1-e_2)^2 = 
 2^8 \pi ^{12} \eta (\tau)^{24}
\eea
where $2\eta ^3 = \tet_2\tet_3\tet_4$ is the Dedekind $\eta$-function.
It remains to compute the combinations $\zeta _{1,3}$ defined in (\ref{periods1}).
The starting point is \cite{magnus},
\bea
\label{zetas}
\zeta  _1 = \zeta _3 + { i \, \pi \over 2 \omega _1 \omega _3} =
  - {i \pi \over  \omega _1^2} \p_\tau \ln \eta (\tau)
\eea
which shows that $\zeta _1$ and its $\tau \to -1/\tau$ transform $\zeta _3$
are modular connections.
Using the relations between $\tet^4$ and $\p_\tau \ln \tet$ in the second 
column of  (\ref{branch1}), we also obtain the following formulas for $i=1,2,3$, 
\bea
e_i + \zeta  _1 = 
- {i \pi \over  \omega _1^2} \p_\tau  \ln \tet_{i+1} (0,\tau)  
\eea
We conclude by noticing that the modular objects $M_{1,2}^{(m)}$ are 
not modular forms, but modular connections, as the parts proportional
to $\zeta _1$ and $\zeta _3$ transform inhomogeneously under the 
modular group $SL(2, \bZ)$.

\subsection{The special case  of the square torus}

For the square torus, $\tau = i$, the above quantities may be
evaluated in elementary terms and the period relations may be solved 
explicitly. Without loss of generality, we choose 
the half-periods in the canonical normalization, $\omega _1 = 1/2$ and 
$\omega _3 = \tau /2$. Using the symmetry of the square torus, 
we have $\zeta _3 = - \zeta _1$. Using the first equation in (\ref{zetas}), 
we find  $\zeta _1 = \pi$ and $\zeta _3 =-\pi$.
In terms of the parameters of the lower half-plane representation, we have
\bea
e_1 = -e_3= k   >0 & \hskip 1in & \, e_2 = 0 
\no \\
E_1 =\, E_3 = 2 k^2 \hskip 0.2in & \hskip 1in &   E_2 = - k^2
\eea
where $k \equiv \wp (1/2)$, whose numerical value is approximately $k=6.875185816$. 
The ranges of the real zeros are as follows,
\bea
\alpha _2 < -k < \beta _2 <  0 < \a_1 < k < \beta _1
\eea
The modular objects $M_1 ^{(m)}$ and $M_2^{(m)}$, defined in 
(\ref{modular12}) take the form,
\bea
2M_1 ^{(2m)} = - \pi  k^{2m-4} \left ( 1  + 2\delta _{m,0} \right )
& \hskip 1in & 
2M_1 ^{(2m+1)} =  k^{2m-2}
\no \\
2M_2 ^{(2m)} = + \pi  k^{2m-4} \left ( 1 + 2 \delta _{m,0} \right )
& \hskip 1in & 
2M_2 ^{(2m+1)} =  k^{2m-2}
\eea
and are used to calculate the parameters $a_n$ and $b_n$ in the period relations 
(\ref{circles}), and we find, 
\bea
2k^4 a_0 & = & - 3 \pi \a_1 \a_2  - (\a_1 + \a_2)k^2 - \pi k^2
\no \\
2k^4 b_0 & = & + 3 \pi \b_1 \b_2  - (\b_1 + \b_2)k^2 + \pi k^2
\no \\
2k^2 a_1 & = & + \a_1 \a_2  + \pi (\a_1 + \a_2) - k^2
\no \\
2k^2 b_1 & = & + \b_1 \b_2 - \pi (\b_1 + \b_2) - k^2
\no \\
2k^2 a_2 & = & - \pi \a_1 \a_2 +  (\a_1 + \a_2)k^2 - 3 \pi k^2
\no \\
2k^2 b_2 & = & + \pi \b_1 \b_2 +  (\b_1 + \b_2)k^2 + 3 \pi k^2
\eea
One could now proceed and impose the conditions (\ref{circles2}) for the 
existence of solutions.

\smallskip

It turns out that a more explicit description of the allowed parameter
space may be obtained by leaving the complex zero $u_1$ as a 
known parameter and solving instead for the real zeros $\alpha _2$ and 
$\beta _1$, as was done also in subsection 5.3.  
It is convenient to scale a factor of $k$ out of $u_1$,
\bea
u_1 = k v
\eea
In terms of $v$ in the lower half-plane, $\Im (v)<0$, the relations (\ref{circles}) are linear in 
$\alpha _2$ and $\b_1$, and are solved as follows,
\bea
\a_2 & = & - { \pi k^2 (|v|^2 +3) + \a_1 k^2 (|v|^2 -1) + (\pi \a_1 - k^2) k (v + \bar v)
\over 
\pi \a_1 ( 3 |v|^2 +1) + k^2 (|v|^2 -1) + (\a_1 + \pi ) k (v + \bar v)}
\no \\
\b _1 & = & { \pi k^2 (|v|^2+3) - \b_2 k^2 (|v|^2 -1) + (\pi \b_2 + k^2) k (v + \bar v)
\over - \pi \b _2 (3 |v|^2 +1) + k^2 (|v|^2 -1) - (\pi - \b_2) k (v + \bar v)}
\eea
These results must be supplemented with the inequalities 
\bea
\label{domainab}
\a_2 & < & -k
\no \\
\beta_1 & > & +k
\eea 
For the special point $v=-i$, it is clear that both inequalities (\ref{domainab}) 
hold for all allowed values $ -k < \b_2 < 0<\a_1<k$. In the next subsection, 
we shall show that the  corresponding solution  may actually be mapped onto 
the Janus solution. 

\smallskip

For non-special points $v \not= -i$, the solution is distinct from Janus. 
Given the ranges $-k < \b_2 <0<\alpha _1  <k$ with strict inequalities,  
it is clear from (\ref{domainab}) that there will exist an open set 
containing $v=-i$ for which the inequalities
will be satisfied, and thus regular solutions will exist.

\subsection{Supersymmetric Janus as a limiting case}

We now return to the general torus in the formulation of (\ref{diffs}) and (\ref{ellhes}).
There is one type of symmetric assignment of the $A_\alpha$ and $B_\alpha$
for which all inequalities and periods relations are automatically satisfied,
and which precisely reproduces the genus 0 Janus solution. The assignment is
given as follows,
\bea
A_0 = - A_2 = -1 & \hskip 1in & B_0 = - B_2  = -1
\no \\
A_1 = -A_3 = -a &\hskip 1in & B_1 = - B_3 = +b 
\eea
where $a,b >0$. In $\p h_1$ and $\p h_2$,  the dependences
on $\zeta$ cancel, and one is left with
\bea
\p h_1 (z) & = & -i \wp (z) +i \wp \left (z + \omega _2 \right ) -ia \wp (z+\omega _1)
+ i a \wp (z+ \omega _3) 
\no \\
\p h_2 (z) & = & - \wp (z) + \wp \left (z + \omega _2 \right ) + b \wp (z+\omega_1)
- b \wp (z+ \omega _3)
\eea
Since $\p h_1 ( \omega _2/2) = \p h_2 (\omega _2/2) = 0$, the point
 $z=\omega _2/2$ is the common complex zero $u_1 = \wp (\omega _2/2)$. 
(For the square torus, and using  formula (15) of section 13.13 of \cite{bateman}, 
we find indeed that $u_1= - i k$.) To identify this solution with Janus, 
we compute the functions $h_1$ and $h_2$ for this assignment, 
with the help of (\ref{ellhes}). Using the formula (\ref{zetadd}),
as well as the fact that $\zeta (\omega _1)$ is real and $\zeta (\omega _3)$ is 
imaginary, we recast the functions as follows, 
\bea
h_1(z) & = & {i \over 2} \wp '(z) \left ( 
- {1 \over \wp (z) - e_2} + {a \over \wp (z) - e_1} - {a \over \wp (z) - e_3} \right ) + {\rm c.c.}
\no \\
h_2(z) & = & {1 \over 2} \wp '(z) \left ( 
- {1 \over \wp (z) - e_2} - {b \over \wp (z) - e_1} + {b \over \wp (z) - e_3} \right ) + {\rm c.c.}
\eea
We now identify with the variables of the Janus solution $u,r$, by identifying the 
functions $h_1$ and $h_2$, which requires the following map,
\bea
\sqrt{u} & = & - {1 \over 2}  {\wp' \over \wp - e_2}
\no \\
{1 \over \sqrt{u}} & = &  - {b \over 2}  {\wp' \over \wp - e_1} + {b \over 2}  {\wp' \over \wp - e_3} 
\no \\
- {r \over \sqrt{u}} & = &  + {a \over 2}  {\wp' \over \wp - e_1}  - {a \over 2}  {\wp' \over \wp - e_3}
\eea
Consistency of these three equations requires first the proportionality of the last
two equations, or, $r =  a / b$, as well as the relation obtained by taking the sidewise 
product of the first two equations. After simplification of the $\wp$-dependence, 
using (\ref{defwp}), we have $1= b (e_1-e_3)$. This is always possible since, 
by our conventions, we have $e_1>e_3$.

\subsection{Homology 3-spheres and 3-form charges}

The homology 3-spheres in this geometry are $S_1^3 = [e_2,e_1]\times _f S_1^2$
and $S_2^3 = [e_3,e_2]\times _f S_2^2$. The corresponding charges are 
\bea
\cH & = & {8 \pi ^2 \over \omega _1} \left ( 1 - \sum _{i=1} ^3 B_i \right )
\no \\
\cC & = & {8 \, i \, \pi ^2 \over \omega _3} 
\left ( 1 - \sum _{i=1} ^3 A_i \right )
\eea
Notice that these vansih on the Janus solution, but are otherwise generally
non-vanishing.

\newpage

\section{Higher genus  solutions}
\setcounter{equation}{0}

At genus higher than 1, the explicit solution involves hyperelliptic integrals,
whose explicit form is less familiar than that of elliptic functions. Nonetheless,
partial analytic and numerical study of the solutions is possible. In subsection \ref{genus2sol}, 
the general set-up needed to obtain regular solutions at genus 2 is presented, 
while in subsection \ref{continuity}  a general continuity argument is given for 
the existence of completely regular solutions, which satsify
all the positivity conditions, at all genera. Finally,  the solution 
may be formulated in terms of higher genus $\tet$-functions, which is done 
in subsection \ref{theta}.

\subsection{The genus 2 solutions}
\label{genus2sol}

We follow closely the construction given in subsections \ref{sectionh1h2}
and \ref{sectionperiodrelations} valid for any genus.
We normalize the genus 2 curve so that
\bea
s(u)^2 & = & (u-e_1)  (u-e_2)(u-e_3)  (u-e_4)(u-e_5)  
\no \\
0 & = & e_1 + e_2 + e_3 + e_4 + e_5
\no \\
2g_2 & = & e_1^2  + e_2 ^2  + e_3^2  + e_4 ^2 + e_5 ^2
\eea
It is straightforward to compute $G_i(u)$, and we find,
\bea
G_i(u) = (u-e_i) \left ( - 3 u^2 - 4 u e_i - 3 e_i ^2 + g_2 \right )
\eea
The  complex zeros $u_1, u_2, \bar u_1, \bar u_2$, with $\Im(u_1), \Im (u_2) <0$,
and the real zeros $\a_1, \a_2, \a_3$ and $\b_1, \b_2, \b_3$, enter as follows,
\bea
P(u) & = & (u-u_1)(u-\bar u_1)(u-u_2)(u-\bar u_2)
\no \\
Q_1 (u) & = & (u-\a_1)(u-\a_2)(u-\a_3)
\no \\
Q_2(u) & = & (u-\b_1)(u-\b_2)(u-\b_3)
\eea
Positivity of $h_1,h_2$ and negativity of $W$ require the following ordering,
\bea
\a_3<e_5<\b_3<e_4<\a_2<e_3<\b_2<e_2<\a_1<e_1<\b_1
\eea
The polynomials $R_1(u)$ and $R_2(u)$ are obtained from the 
expansions at $u=\infty$, and we find,
\bea
R_1 (u) & = & u^2 - A_7 u + A_6
\no \\
R_2(u) & = & u^2 - B_7 u + B_6
\eea
where the coefficients are given by
\bea
A_7 & = & 2 \Re(u_1+u_2) + \a_1 + \a_2+ \a_3
\no \\
B_7 & = & 2 \Re(u_1+u_2) + \b_1 + \b_2+ \b_3
\no \\
A_6 & = & |u_1|^2 + |u_2|^2 + 4 \Re(u_1)\, \Re(u_2)
	+ 2 \Re(u_1+u_2)(\a_1 + \a_2+ \a_3)
	\no \\ && \quad
	+ g_2 + \a_1 \a_2+ \a_2 \a_3+ \a_3 \a_1
\no \\
B_6 & = & |u_1|^2 + |u_2|^2 + 4 \Re(u_1)\, \Re(u_2)
	+ 2 \Re(u_1+u_2)(\b_1 + \b_2+ \b_3)
	\no \\ && \quad
	+ g_2 + \b_1 \b_2+ \b_2 \b_3+ \b_3 \b_1
\eea
Putting all together, the period relations are as follows, 
\bea
\label{genus2periods}
k=1,2 & \hskip 0.7in & \int _{e_{2k}} ^{e_{2k-1}} {du \over s(u)}
\bigg [ u^2 - A_7 u + A_6 + \sum _{i=1}^5 A_i G_i(u) \bigg ]=0
\no \\
k=1,2 &&  \int _{e_{2k+1}} ^{e_{2k}} {du \over s(u)}
\bigg [ u^2 - B_7 u + B_6 + \sum _{i=1}^5 B_i G_i(u) \bigg ]=0
\eea
where
\bea
A_i & = &   - P(e_i) Q_1(e_i) \, (E_i)^{-2}
\no \\
B_i & = &   - P(e_i) Q_2(e_i) \, (E_i)^{-2}
\eea
The four period integrals above take on either real or purely imaginary
values, as a result of the phase values of (\ref{sphase}) for $s(u)$.
Since the $s(u)$ retains constant phase inside any interval between
consecutive branch points, the denominator 
$s(u)$ in the period integrals above may be replaced by $|s(u)|$. 
Given the moduli of the surface, and the real zeros $\a_1,\a_2,\a_3$ and 
$\b_1,\b_2,\b_3$, these 4 real equations then determine the two complex
zeros $u_1$ and $u_2$.

\subsection{Numerical analysis}

We have explored the solutions of these equations numerically for genus 2 
surfaces with various degrees of $\bZ_2$ symmetry, obtained via the following 
arrangement of  the branch points,
\bea
e_5 = - t_2, \qquad e_4 = -{1 \over t_2}, \qquad e_3 =0, \qquad 
e_2 = {1 \over t_1}, \qquad e_1 = t_1
\eea
for $t_1, t_2 >1$. The first $\bZ_2$ is the hyperelliptic involution, while a 
second  $\bZ_2$ is the inversion $u \to 1/u$. An extra $\bZ_2$, given by 
$u \to -u$, is  obtained by setting $t_2=t_1$, while a Riemann surface with 
symmetry $(\bZ_2)^4$ is obtained when $t_1=t_2 = t_0$, where $t_0 \equiv 1.27201965$. The  period matrix  for the surface with $(\bZ_2)^4$ symmetry 
assumes the symmetrical form,
\bea
\Omega _{IJ} = \left ( \matrix{0.89442719 \, i & & - 0.44721359 \, i \cr
0.44721359 \, i && 0.89442719 \, i\cr} \right )
\qquad I,J=1,2
\eea

\begin{figure}[tb]
\begin{center}
\epsfxsize=4in
\epsfysize=3in
\epsffile{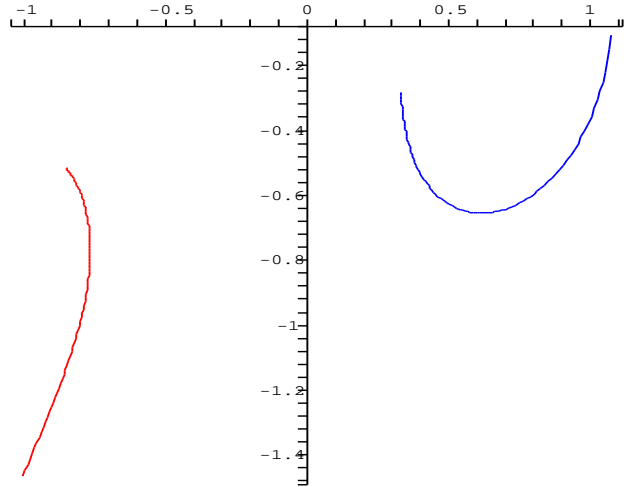}
\label{figure9}
\caption{Genus 2 solutions for complex zeros $u_1$ (red) and $u_2$ (blue)
in the lower half $u$-plane, for modulus $t_1$ in the interval $1.03<t_1< 2$, 
fixed modulus $t_2=t_0$, and $\a_1=(e_1+e_2)/2$, $\a_2 = (e_3+e_4)/2$,
$\a_3 = 2e_5$, $\b_1 = 2e_1$, $\b_2 = (e_2+e_3)/2$, and $\b_3 = (e_4+e_5)/2$. }
\end{center}
\end{figure}

In Figure 7, we make a choice for the real zeros indicated on the figure's caption, 
and plot the complex
zeros $u_1$ and $u_2$, obtained as solutions to the period relations
(\ref{genus2periods}), as a function of the modulus $t_1$. The existence
of this numerical solution means that global regular solutions exist
for $h_1$ and $h_2$ for the above assigned values. In addition, we have 
numerically shown that regular solutions exist for less symmetrical assignments
of the real zeros, and modulus $t_2 \not= t_0$.
It is an interesting problem to describe the region of parameters
$\a_1, \a_2, \a_3, \b_1, \b_2, \b_3$ and moduli, but this issue lies
beyond the scope of this paper.

\smallskip

Note that the complex zero $u_2$ tends to the real axis as $t_1 \to 1$.
This limiting case corresponds to a degeneration of the genus 2 surface,
a phenomenon that will be analyzed in generality in the subsequent section.

\subsection{Parametrization via $\tet$-functions}
\label{theta}

The Thomae formulas \cite{mumford} allow us to express the branch points in 
terms of genus two  $\tet$-functions. We begin by recalling some key facts 
about genus two $\tet$-functions \cite{mumford,DPIV}.
A general spin structure is a half-interger characteristic, consisting of an array of two 2-component vectors,
\bea
\kappa = \left ( \kappa ' | \kappa '' \right ) 
\hskip 1in  \kappa ' = \left ( 
\matrix{ \kappa '_1 \cr \kappa '_2 \cr} \right )
 \hskip 1in  \kappa '' = \left ( 
\matrix{ \kappa ''_1 \cr \kappa ''_2 \cr} \right )
\eea
and the entries $\kappa '_{1,2}$ and $\kappa '' _{1,2}$ take on values $0$ or 
$1/2$ mod 1.
The $\tet$-function is defined in terms of the period matrix $\Omega$, and a
general complex 2-component vector $\zeta$ by
\bea
\tet [\kappa] (\zeta, \Omega) 
=
\sum _{n \in \bZ^2} \exp \left \{ \pi i (n+\kappa ')^t \Omega (n + \kappa ')
+ 2 \pi i (n+\kappa ')^t(\zeta + \kappa '') \right \}
\eea
For any spin structure $\kappa$, the combination $ 4 (\kappa ')^t \kappa ''$ is an integer:
even (resp. odd) spin structures correspond to $ 4 (\kappa ')^t \kappa ''$ even (resp. odd).

\smallskip

For genus 2, there exists a sharper version of the Thomae
formulas, which was heavily used in \cite{DPIV}. The key is a one-to-one map
between the six branch points $e_i$, $i=1,\cdots, 6$ with $e_6 =\infty$, and 
the six odd spin structures $\nu _i$ at genus 2. The relation with $\tet$-functions 
is then simply given by
\bea
\label{crossratio}
{(e_i - e_j)(e_k - e_l) \over (e_i-e_k)(e_j-e_l)}
= {\cM _{\nu _i \nu _j} \cM _{\nu _k \nu _l} \over \cM _{\nu _i \nu _k}
\cM _{\nu _j \nu _l} }
\eea
where $\cM _{\nu _i \nu _j}$ is a modular form of weight 2, defined by
\bea
\cM _{\nu _i \nu _j} = 
\p_1 \tet [\nu _i] (0, \Omega ) \p_2 \tet [\nu_j] (0,\Omega)
- \p_2 \tet [\nu _i] (0, \Omega ) \p_1 \tet [\nu_j] (0,\Omega)
\eea
An alternative formula for $\cM _{\nu _i \nu _j}$, with $i\not= j$, is in terms of even spin
structures, but is given only up to an overall sign (familiar from the standard
Thomae formulas of \cite{mumford}),
\bea
\cM _{\nu _i \nu _j}^2 
= \pi ^4 \prod _{k \not= i,j} \tet [\nu_i + \nu_j + \nu_k](0,\Omega )^2
\eea
Upon choosing $e_4=0$, $e_5 =1$, and $e_6=\infty$, by $SL(2,\bR)$ invariance, 
and $i=4$, $k=5$ and $l=6$, the cross ratio formula yields an explicit formula
for the 3 remaining real moduli, 
\bea
e_j
=  {\cM _{\nu _j \nu _4} \cM _{\nu _5 \nu _6} \over \cM _{\nu _j \nu _6} \cM _{\nu _5 \nu _4}}
\eea
Note that this formula actually holds for all possible values of $j=1,\cdots, 6$.

\smallskip

The period integrals that enter into the period relations (\ref{genus2periods})
are half $\cA$- and $\cB$-cycle integrals of Abelian differentials. We shall declare 
the following correspondence,
\bea
[e_2, e_1]  \sim  \cA_1 & \hskip 1in & [e_4, e_3] \sim \cA_2
\no \\
{} [e_3, e_2] \sim \cB_1 & & [e_5, e_4] \sim \cB_2
\eea
We define the following periods for $k=0,1,2,3$, 
\bea
\int _{e_2} ^{e_1} {u^k du \over s(u)} = K_{1,k} 
& \hskip 1in & 
\int _{e_4} ^{e_3} {u^k du \over s(u)} = K_{2,k}
\no \\
\int _{e_3} ^{e_2} {u^k du \over s(u)} = L_{1,k} 
& \hskip 1in & 
\int _{e_5} ^{e_4} {u^k du \over s(u)} = L_{2,k}
\eea
The periods  $L$ for $k=0,1$ are given in terms of the period matrix $\Omega$
and the periods $K$ by the relations, 
\bea
\Omega  = {1 \over K_{1,1} K_{2,0} - K_{1,0} K_{2,1}}
\left ( \matrix{ 
K_{2,0} L_{1,1} - K_{2,1} L_{1,0} && K_{2,0} L_{2,1} - K_{2,1} L_{2,0} \cr
& \cr 
K_{1,1} L_{1,0} - K_{1,0} L_{1,1} && K_{1,1} L_{2,0} - K_{1,0} L_{2,1} \cr} \right )
\eea
The $K$ periods cancel out of the set of four period relations (\ref{genus2periods}),
just as an overall factor of $\omega _1$ cancelled out of the genus 1 period relations.

\smallskip

The periods $K$ and $L$ for $k=2,3$ are analogous to the quantities $\zeta (\omega_1)$
and $\zeta (\omega _3)$ encountered at genus 1. They are the periods of 
Abelian differentials with poles of order 2 and 4 at the branch point $e_6 =\infty$.
We are aware of no known explicit formulas
for these periods in terms of $\tet$-functions, but suspect that such formulas
could be derived with the help of the conversion formulas between the 
$\tet$-function and hyperelliptic representations obtained in \cite{DPIV}.

\smallskip

The differentials $\p h_1$, $\p h_2$ as well harmonic functions may also
be expressed in terms $\tet$-functions. The key ingredient is the identification of 
the Abelian differentials with double poles at the branch points with their form 
in terms of $\tet$-functions, a result available through the use of the prime form
\cite{Fay}. Since the definition of the prime form involves the Abelian 
integrals of the first kind anyway, it is unclear how much would be gained from
this alternative expression, and we shall therefore suppress these formulas.

\subsection{Continuity argument for existence of solutions at all genera}
\label{continuity}

The solutions for different genera form a connected set and the lower 
genus solutions may be obtained as a limit of higher genus solutions, by
collapsing branch cuts. For example, a genus $g$ solutions, with branch 
points $e_1 , \cdots , e_{2g+1}$ and $e_{2g+2}=\infty$, real zeros 
$\a_1, \cdots \a_{g+1}$, and $\b_1 \cdots , \b_{g+1}$, and complex
zeros $u_1 , \cdots , u_g$, is smoothly connected to the genus $g-1$
solution by letting 
\bea
e_{2b}, \, e_{2b-1} & \to & x
\no \\
\a_b , \b _b , u_b , \bar u_b & \to & x
\eea
for some $1 \leq b \leq g$.
In this limit, the resulting differentials $\p h_1$ and $\p h_2$ become independent
of the point $x$, and collapse onto the genus $g-1$ solution, obtained from the
above genus $g$ solution with the branch points $e_{2b}, e_{2b-1}$ and the zeros
$\a_b, \b_b, u_b, \bar u_b$ removed. This limit is completely smooth. 
Importantly, the limit is a {\sl local} process, which is largely insensitive 
to the global properties of $\Sigma$.

\smallskip

Now consider the reverse problem. If a genus $g-1$ solution
exists, the question is naturally raised as to whether a genus $g$ solution
exists {\sl in an open neighborhood of parameter space of the genus $g-1$ solution}. 
In other words, can an extra genus be ``turned on" by inserting 2 extra branch points 
$e_{2g+1}, e_{2g+2}$  and adding zeros $\a_g, \b_g, u_g, \bar u_g$? We have 
investigated this question numerically for genus 1 and 2, and have found
an affirmative answer to this question. We believe that it should be possible
to prove analytically that a genus $g$ solution will indeed exist in a sufficiently small open
neighborhood of any genus $g-1$ solution. 

\smallskip

One key ingredient in the above conjecture is the necessary condition that, as a 
branch cut collapses to a point $x$, exactly one complex zero converges to
the real axis and collapses to $x$ as well. This collapse phenomenon is 
studied in the next section, where it is shown to hold in general. It is also
shown there, that, if one allows for mild singularities in the 10-dimensional 
geometry, such as those produced by D5 and NS5 branes in the probe limit, 
then  only one of the real zeros, either $\a$ or $\b$ is required
to collapse to $x$ as well, but not both. Thus, allowing for such probe limit
singularities,  a higher-dimensional parameter space of  degenerations is allowed.

\newpage

\section{Collapse of branch cuts and D- and NS-branes}
\setcounter{equation}{0}

A direct study of the collapse of branch cuts from the explicit hyper-elliptic 
solution is complicated by the fact that the splitting off of pole terms in $h_1$ and 
$h_2$, between the functions $p_{1,2}$ and $q_{1,2}$ as in formula (\ref{h1h2})
introduces artificial divergences in $p_{1,2}$ and $q_{1,2}$ which however 
cancel in the functions $h_1$ and $h_2$. For this reason, we shall take first the 
limit of collapsing branch cuts on the differentials $\p h_1$ and $\p h_2$,
where this limit exists consistently, and then re-examine the questions of regularity.

\subsection{The case of genus 1}

For  genus 1, the branch points and zeros of $Q_1$ and $Q_2$
are subject to the following ordering,
$\a_2 < e_3 < \b_2 < e_2 < \a_1 < e_1 < \b_1$.
We shall take the limit where $e_1 - e_2 \to 0$, which clearly forces also 
$e_1 - \a_1 \to 0$. Without loss of generality, we  make an overall translation
by $e_3$, and set $e_3=0$, which is the position of the only remaining branch point, and we use the designation $e_1, e_2, \a_1 \to k^2>0$, with $k>0$, and $\a=\a_2$.  Thus, we have the ordering
\bea
\label{order2}
\a < 0 < \b_2 < k^2 < \b_1
\eea
and the differentials are 
\bea
\p  h_1  & = & -i \, {(u-u_1)(u-\bar u_1) (u - \a)
\over (u-k^2)^2}  \, { du \over \sqrt{u}^3}
\no \\
\p  h_2  & = & -  \, {(u-u_1)(u-\bar u_1) (u - \b_1)(u-\b_2)
\over (u-k^2)^3} \, { du \over \sqrt{u}^3}
\eea
As a smooth limit of the elliptic case with the ordering 
of $\a, k^2, \b_1,\b_2$ prescribed above, the regularity condition $W\leq 0$ is automatic. 
To uniformize the square root, we introduce a new 
coordinate $u= w^2$, which, in view of $\Im (u)<0$,  takes values in the second quadrant,
\bea
\label{domain}
\Sigma \equiv \left \{ w; \, \Re (w) <0, \,  \Im (w) >0 \right \}
\eea
To work out the vanishing Dirichlet conditions on $h_1$ and $h_2$, and insist on their 
positivity, we decompose the fractions into elementary poles, and we have,
\bea
\p  h_1  & = & -2i
\left [  1 + {A_1 \over w^2} + {2k B_1 \over w^2-k^2} + 
{2C_1(w^2+k^2) \over (w^2-k^2)^2}  \right ] dw
\\
\p  h_2  & = & - 2
\bigg [ 1 + {A_2 \over w^2} + {2k B_2 \over w^2-k^2}  + 
{2 C_2 (w^2+k^2) \over (w^2-k^2)^2}  
+ {2D_2 (3w^2k +k^3)\over (w^2-k^2)^3}  \bigg ]dw
\no
\eea
It is straightforward to integrate these (up to an additive constant, which must vanish by the vanishing Dirichlet conditions),
\bea
h_1 & = & -2i (w-\bar w) \left [ 1 +{A_1 \over |w|^2} + {2 C_1 (|w|^2 + k^2) \over |w^2-k^2|^2} 
 \right ]
 -2i B_1 \ln  { (w-k)(\bar w +k) \over (w+k) (\bar w -k)} 
\no \\
h_2 & = & -2 (w+\bar w) \left [ 1 - {A_2 \over |w|^2} + 4 C_2 { |w|^2 - k^2 \over |w^2-k^2|^2}
+ 4 k D_2 {|w|^2(w^2+\bar w^2 - w\bar w - 2 k^2) + k^4 \over |w^2 - k^2|^3} \right ]
\no \\ && \hskip 1in - 2 B_2 \ln {|w-k|^2 \over |w+k|^2}
\eea
For $w \in \Sigma$, we have $-2i(w-\bar w) >0$ and $-2 (w + \bar w)>0$. 
Therefore, positivity of  $h_1$ and $h_2$  at the poles $w=0,-k$
(note that $w=+k \not \in \Sigma$) requires that 
\bea
\label{ineq1}
A_1 >0 & \quad B_1 =0  \quad & C_1 \geq 0
\no \\
A_2 <0 & \quad k B_2 \leq 0 \quad & C_2 = D_2 =0
\eea
The requirements on $A_1, A_2, C_1$ are obvious. The requirement $B_1=0$
stems from the fact that as $w$ crosses the point $w=-k$, the logarithms 
pick up an additive contribution of $2 \pi i$, so that the left and the right 
Dirichlet boundary values of $h_1$ differ by $2 \pi B_1$. But regularity of the 
solution requires that all Dirichlet boundary values vanish, so that we must have $B_1=0$.
Finally, it is easy to see that the sign of the numerator of the $D_2$ term depends
on how the point $w=-k$ is being approached. Setting $\Re(w)=-k$ and varying its
imaginary part produces a negative numerator, while setting $\Im (w)=0$
and varying its real part produces a positive numerator. Since this is the leading 
singularity at $w=-k$, we must have $D_2=0$. Given $D_2=0$, the term in  
$C_2$ remains
as the leading singularity and the same argument applies to conclude that $C_2=0$, which concludes the proof of (\ref{ineq1}).

\begin{figure}[tb]
\begin{center}
\epsfxsize=4in
\epsfysize=3.5in
\epsffile{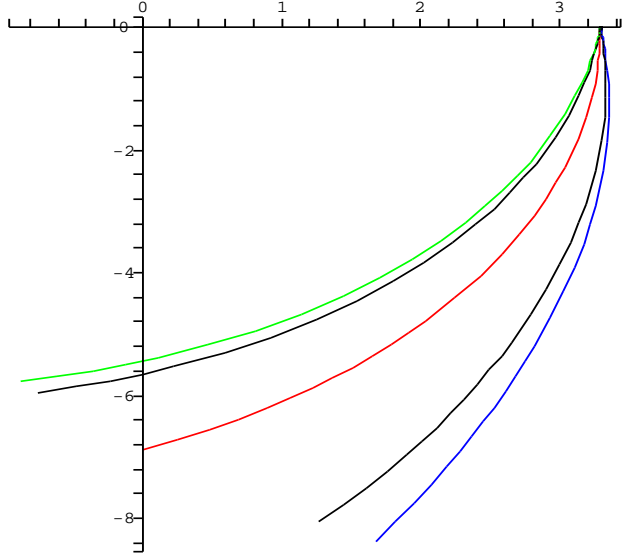}
\label{figure6}
\caption{Genus 1 collapsing branch cut, starting with a square torus at the top of
each curve, to $e_1-e_2 \to 0$ at the real axis, for $\a_1 = (e_1+e_2)/2$, $\a_2 = 2 e_3$, 
$\beta _1 = 2 e_1$ and various values of $\beta _2 = e_3 + x (e_2-e_3)$, 
given by $x=0.99$ (blue), $x=0.90$ (top black), $x=0.50$ (red), $x=0.10$ (bottom black),
and $x=0.01$ (green).  }
\end{center}
\end{figure}

Obtaining the values of $A_1,B_1, C_1, A_2, B_2, C_2, D_2$ from those of $k, \a, \b_1,
\b_2$, we have 
\bea
A_1 = - \a { |u_1|^2 \over k^4} \hskip 0.7in &  & 
A_1 + 2 C_1 = - u_1 - \bar u_1 - \a + 2k^2
\no \\
A_2 =  - {\b_1 \b_2 |u_1|^2 \over k^6} \hskip 0.5in & & 
A_2 + 2 k B_2 = - u_1 - \bar u_1 - \b_1 - \b_2 + 3 k^2
\no \\
C_1 = {|k^2 - u_1|^2 (k^2-\a)\over 4 k^4} & \hskip 0.5in & B_1=0
\eea
together with the requirement that $(u-u_1)(u-\bar u_1) (u - \b_1)(u-\b_2)$ 
have a double zero in $u$ at $u=k^2$, which is equivalent to $C_2=D_2=0$.
Note that the conditions $A_1>0, A_2 <0, C_1>0$ of (\ref{ineq1}) follow
directly from the above identifications and the ordering (\ref{order2}).

\medskip

To solve the remaining conditions, it is helpful to first solve the $C_2=D_2=0$
condition. There are two cases. Either the double zero of
 $(u-u_1)(u-\bar u_1) (u - \b_1)(u-\b_2)$  is simply $u_1=k^2$, or it is not: $u_1 \not= k^2$.

\subsubsection{The case $u_1 \not= k^2$}

In this case, the double zero requires $\b_1=\b_2=k^2$, and we have 
$2k^3 B_2 =  | k^2-u_1|^2 >0$,
and hence the condition $k B_2\leq 0$ can never be satisfied. Thus, this case
is simply ruled out.

\subsubsection{The case $u_1 = k^2$}

This leads to drastic simplifications, and we have
\bea
A_1 = - \a \hskip 0.3in & \hskip 0.5in & B_1 = C_1=C_2=D_2=0
\no \\
A_2 = - {\b_1 \b_2 \over k^2} && B_2 = {1 \over 2 k^3} (k^2-\b_1)(k^2-\b_2)
\eea
We see that the condition $k B_2 \leq 0$ is now automatic as well.
The solution may be written down explicitly. The harmonic functions are given by
\bea 
\label{h12formula}
h_1 & = & - 2i (w-\bar w) \left [ 1 - { \a \over |w|^2} \right ]
\no \\
h_2 & = & - 2 (w + \bar w) \left [ 1 + {\b_1 \b_2 \over k^2 |w|^2} \right ]
+ {(\b_1 - k^2) (k^2 - \b_2) \over k^3} \, \ln {|w-k|^2 \over |w+k|^2}
\eea
Since $w \in \Sigma$ of (\ref{domain}), and using the ordering (\ref{order2}), 
it is immediate that
$h_1, h_2>0$ everywhere in $w\in \Sigma$. In particular, the $\ln$ is always 
positive since for $w \in \Sigma$, we have $|w-k|^2 > |w+k|^2$. Notice that, 
if we also have $\beta_1=k^2$ or $\beta _2=k^2$, then we recover the Janus solution.

\subsection{General collapse of a branch cut}

Having derived the behavior under collapse of a branch cut in the genus 1 case,
we are now in a position to generalize the result to the collapse of a branch cut
for general genus. The reason that such a result can be obtained is that the 
regularity and positivity conditions are essentially local conditions,
largely insensitive to the global properties of the surface $\Sigma$.

\smallskip

Below, we shall show the following general result.  {\sl The collapse of any 
branch cut is accompanied by the convergence of precisely one of the 
complex zeros $u_a$ to the real axis, and more specifically, to the location 
of the collapsing branch cut.}

\smallskip

For simplicity, we shall assume that two consecutive branch points $e_{2b}$
and $e_{2b-1}$ collapse to a zero of type $\alpha$, and that the adjacent 
branch points $e_{2b+1}$ and $e_{2b-2}$ remain a finite distance away 
from $\a$. (If  adjacent branch points are allowed to collapse onto $\a$ as 
well, we are dealing with a multiple degeneration of the Riemann surface. 
The same methods, to be explained below, can also
be adapted to cover those cases.)

\smallskip

The corresponding branch cut $[e_{2b} , e_{2b-1}]$ contains a single  zero 
$\a_b$ of $Q_1$.
(The argument may readily be carried over to an interval $[e_{2b+1}, e_{2b}]$
which would contain a single zero $\b_{b+1}$ of $Q_2$ instead.)
Since $\a_b \in [e_{2b} , e_{2b-1}]$, this means that also $\a_b \to \a$. 
The resulting ordering of the collapsed interval is thus,
\bea
\a_{g+1} < e_{2g+1} <  \cdots 
< e_{2b+1} < \b _{b+1} < \a < \b _b < e_{2b-2} < 
\cdots < e_1 <\b_1
\eea
Having taken the limit $e_{2b}, e_{2b-1}, \a_b \to \a$, we shall now study 
the problem locally around $u \sim \a$. To do so, it is convenient to define the following quantities,
\bea
s(u)^2 & = & (u - \a)^2 \tilde s(u)^2 
\no \\
Q_1 (u) & = & (u - \alpha) \tilde Q_1 (u)
\eea
where
\bea
\tilde s(u)^2 & = & ( u - e_{2g+1} ) \prod _{j=1, \,   j \not= b}^g  (u - e_{2j-1} )(u-e_{2j})
\no \\
\tilde Q_1(u) & = & \prod _{a=1, \,  a\not=b} ^g (u - \a _a)
\eea
and recast the differentials $\p h_1$ and $\p h_2$ in terms of the following functions,
\bea
\p h_1 = -i M_1 (u) { du \over (u - \a)^2} 
& \hskip 1in & 
M_1 (u)  \equiv  { P(u) \tilde Q_1 (u) \over \tilde s (u)^3}
\no \\
\p h_2 = - M_2 (u) { du \over (u - \a)^3} 
& \hskip 1in & 
M_2 (u)  \equiv  { P(u)  Q_2 (u) \over \tilde s (u)^3}
\eea
Note that, because of the extra zero in $Q_1$ at $u = \alpha$, the differential 
$\p h_1$ has a pole of order 2 while $\p h_2$ has a pole of order 3. 
Also, it is readily verified that, since $e_{2b+1} < \a < e_{2b-2}$, we have 
$\tilde s(\a)^2>0$, so that $\tilde s(u)$, and thus $M_1(u)$ and $M_2(u)$ 
are  real for $u$ real and in the neighborhood of $\alpha$. Expanding the 
differentials in powers of $w$, with  $u = \a+w$,
and integrating to obtain the harmonic functions $h_1$ and $h_2$, we find,
\bea
h_1 & = &   
M_1 (\a) {2 \Im (w) \over |w|^2} - i M_1 '(\a) \ln {w \over \bar w} 
+{\rm regular}
\no \\
h_2 & = & 
M_2 (\a) { \Re(w)^2 - \Im (w)^2 \over |w|^4} + M_2 '(\a) { 2 \Re(w) \over |w|^2}
- M_2 '' (\a) \ln |w| + {\rm regular}
\eea
Positivity of $h_2$ clearly requires that
\bea
\label{M2}
M_2 (\a) = M_2'(\a) =0 \hskip 1in M_2''(\a) \geq 0
\eea 
Positivity at the leading singularity of $h_1$ requires that (recall that $u$ and thus $w$
are in the lower half-plane, with $\Im (w) <0$), 
\bea
\label{M1a}
M_1 (\a) \leq 0
\eea
There is also a constraint on the $M_1'(\a)$ term which is more subtle. The function 
$h_1$ must obey vanishing Dirichlet boundary conditions on both sides of $\a$.
The term proportional to $M_1'(\a)$ introduces a discontinuity in $h_1$ along
the boundary, adding to $h_1$ the quantity $2 \pi M_1'(\a)$ as $w$ moves across 
$\a$ along the real axis. No such discontinuity is allowed with vanishing 
Dirichlet boundary conditions, and no such discontinuity can be compensated
for by the regular terms, which have not been explicitly exhibited. Therefore, 
we have the final condition that
\bea
\label{M1b}
M_1 ' (\a) =0
\eea
Clearly, condition (\ref{M2}) requires that $M_2(u)$ must have a double zero at $u=\a$.
There are only two different ways of achieving a double zero. (It is here that we use
the extra assumption, stated above,  that the adjacent branch points
$e_{2b+1}$ and $e_{2b-2}$ remain a finite distance away from $\a$.)

\subsubsection{The case $P(\a)\not=0$}

In this case, we must have $Q_2(\a)=0$, which requires that 
$\b _b = \b _{b+1}=\a$.
We shall now calculate the sign of $M_2''(\a)= P(\a) Q_2 ''(\a) /\tilde s(\a)^3$. 
Analyzing the position of $\a$
relative to the branch points and zeros of $Q_2''$, we readily find,
\bea
{\rm sign} (\tilde s(\a ) )& = & (-1)^b
\no \\
{\rm sign} ( Q_2''(\a )) & = & (-1)^{b-1}
\eea
It follows that $M_2''(\a) < 0$; the  inequality is strict because we have $P(\a)>0$,
as well as $Q_2 ''(\a), \tilde s(\a)  \not= 0$. But this result is contradictory to 
the third result in (\ref{M2}). Therefore, the case $P(\a)\not=0$ is ruled out, and we
must have instead,

\subsubsection{The case $P(\a)=0$}

Since the zeros of $P(u)$ come in complex conjugate pairs, a real zero $\a$
must be a double zero of $P$. It then follows that the first and second 
conditions of (\ref{M2}),
as well as (\ref{M1a}) and (\ref{M1b}) hold automatically. Since we now have 
${\rm sign}(Q_2(\a))= (-1)^b$ and $P''(\a)>0$, it is easy to see that the third
condition in (\ref{M2}) is also satisfied.

\subsection{Complete collapse of the higher genus case}

When branch cuts are being collapsed at higher genus, there are a priori 
two options for the remaining branch point : it can have even or odd index.
If the index is even, $e_{2r}$, for $1\leq r  \leq g$, then either the branch point
$e_1$ or the branch point $e_{2g+1}$ should be collapsed onto the branch 
point at $\infty$ in order to obtain maximal collapse. If the index is odd, $e_{2r-1}$,
$1\leq r \leq g+1$, then the point $\infty$ remains a branch point. 
It is the latter case we shall discuss in detail here, the other case is analogous.
As before, we shift and set $e_{2r-1}=0$. It is clear that the branch cuts
to the left of $0$ collapse with a $\beta$-zero, while the ones to the right 
collapse with an $\alpha$-zero. Following the arguments of subsection 7.2, 
each of the $g$ complex zeros must
collapse to one of the $g$ collapsing branch cuts. We shall denote the negative
collapsed branch cuts $-l_j^2$ and the positive collapsed branch cuts by $+k_i^2$,
with $i=1,\cdots, m= r-1$ and $j=1, \cdots , n=g-r+1$. In view of the above 
considerations, the form of the differentials reduces to
\bea
\p  h_1  & = & - 2 i dw \left [ 
1 + {C_0 \over w^2} + \sum _{j=1}^n { C_j \over w^2 + l_j^2} \right ]
\no \\
\p  h_2  & = & - 2 dw \left [ 
1 + {D_0 \over w^2} + \sum _{i=1}^m { D_i \over w^2 - k_i^2} \right ]
\eea
for  $k_i, l_j >0$. It is straightforward to compute $W$, and check that the condition 
$W<0$ is fulfilled as soon as we have  $C_0 - D_0 >0$, and 
\bea
\label{CD}
C_j >0 &&  j=1,\cdots, m
\no \\
D_i <0 && i=1,\cdots, n
\eea
The functions are also readily evaluated,
\bea
h_1 & = & - 2i (w-\bar w) \left [ 1 + {C_0 \over |w|^2} \right ]
+ \sum _{j=1}^n {C_j \over l_j} \ln {|w+i l_j|^2 \over |w- i l_j|^2}
\no \\
h_2 & = & - 2 (w+\bar w) \left [ 1 - {D_0 \over |w|^2} \right ]
- \sum _{i=1}^m {D_i \over k_i} \ln {|w- k_i |^2 \over |w+ k_i |^2}
\eea
It is immediate that these functions satisfy vanishing Dirichlet boundary conditions,
$h_1=0$ whenever $w$ is real, and $h_2=0$ whenever $w$ is purely imaginary.
The conditions $h_1, h_2 >0$  require  the stronger inequalities $C_0>0$ and $D_0<0$.
For $w \in \Sigma$, using the inequalities, $ |w+i l_j|^2 >  |w- i l_j|^2$ and 
$|w- k_i |^2 > |w+ k_i |^2$,
and the fact that $C_j >0$ while $D_i<0$, as already assumed in (\ref{CD}), it is 
immediate that $h_1, h_2>0$. This completes our proof of regularity of the 
solutions with poles on the boundary of the Riemann surface $\Sigma$.

\subsection{The probe limit}\label{problim}

In this subsection, the behavior of the solutions with collapsed branch points will be analyzed 
near the pole $w=-k$. Note that the two branch points collapse at a zero $\alpha_b$ of 
$\partial h_{1}$.  The goal is to identify the singular solution in this limit with the 
metric of the probe NS5-brane.  The limit when two branch points collapse near 
a zero $\beta_b$ of $\partial h_{2}$ corresponds to a probe D5-brane.

\begin{figure}[tbph]
\begin{center}
\epsfxsize=4.5in
\epsfysize=2in
\epsffile{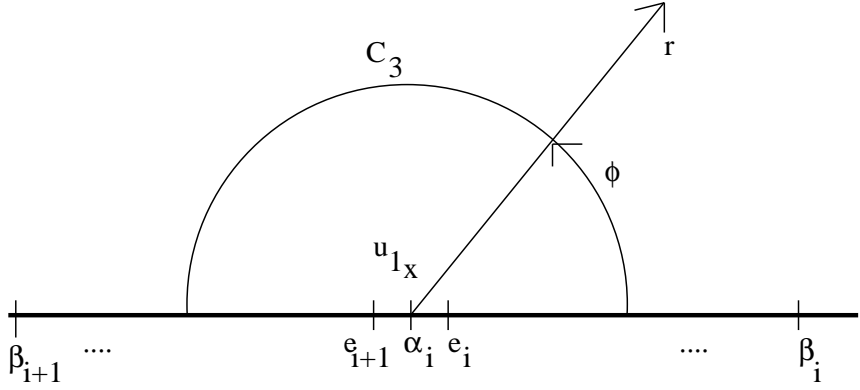}
\label{figure7}
\caption{Local coordinates near two collapsed branch points.}
\end{center}
\end{figure}

We define $w=-k+z$  and introduce a new coordinate $z=re^{i \psi}$.  The expansion 
near the collapsed branch points is around $r=0$. The harmonic functions 
(\ref{h12formula}) have the  expansion
\bea
h_1&=& 4  c_1 r \sin \psi  \nonumber \\
h_2&=& 2 d_0 -2 d_1 \ln r^2
\eea
where $d_0,d_1,c_1>0$ and subleading terms in $r$ and $\ln r $ are dropped. 
In this limit the solution becomes
\bea
\label{limsolz}
e^{4\phi} &=& {d_1^2\over c_1^2} {|\ln (r)|\over r^2}+ \cdots 
\no \\
\rho^2 &= &4 \sqrt{c_1 d_1}  \; {1\over r^{3\over 2} |\ln (r)|^{1\over 4}}+\cdots 
\no \\
f_1^2 &=&  4 \sqrt{c_1 d_1} \; (\sin \psi )^2  {r^{1\over 2} \over |\ln( r)|^{1\over 4} } +\cdots 
\no \\
f_2^2 &=&  4 \sqrt{c_1 d_1} \;  r^{1\over 2} |\ln (r)|^{1\over 4} +\cdots 
\no \\
f_4^2 &=&  4 \sqrt{c_1 d_1} \; r^{1\over 2} |\ln (r)|^{1\over 4}+\cdots 
\no \\
\eea
Note that the range of the angular parameter is $\psi\in[0,\pi]$.  The behavior of 
the metric factors $f_{1}$ and $f_{2}$ given in (\ref{limsolz}) implies that for fixed 
$r$ the  compact  part of the metric is  $S^3 \times S^2$. 
The leading terms for the 2-form potential $B_{(2)}$ are given by 
\bea
b_1&=&  -8 d_1  \psi + 2 \sin 2\psi + o \left ( {1\over |\ln(r)|} \right ) 
\no\\ 
b_2&=& 16 c_1 r \cos \psi + o \left ( {r \over |\ln(r)|} \right ) 
\eea
The  function $b_{1}$ is associated with the NSNS  two form potential and remains 
finite as $r\to 0$. This implies that there is a non-zero NSNS flux through the three 
sphere parameterized by $\psi$ and $S^2_1$. The flux is given by
\bea
\lim_{r\to 0}\int_{S^3}H_3 & =&  
\lim_{r\to  0} \int_{S^3} d b_{1} \wedge \hat e^{45}  
\no \\
&=&  \lim_{r\to 0} \, 4\pi \left  ( b_{1} \bigg |_{\psi=\pi} -  b_{1} \bigg |_{\psi=0} \right )  
\no \\
&=&-32 \pi^{2} d_{1}
\eea
Clearly, the limit of collapsing branch points is singular.  There is, however, 
a natural  interpretation of this singular behavior. The fact that the dilaton 
goes to infinity, that  there is a non-vanishing NSNS 3-form flux through an 
$S^3$ and that metric factors for the $S_2^2$ and $AdS_{4}$ behave 
in the same way all indicate that the singularity can be attributed to the 
presence of a NS5-brane source with worldvolume $AdS_{4} \times S_2^2$. 
At this point, only the flat space NS-brane metric is known. 
A construction of a solution of the supergravity equations in this background 
with an NS5-brane source and a detailed comparison of the solutions is 
beyond the scope of this paper.
   
\smallskip

A similar analysis can be carried out for the collapse of two branch points near 
a zero of $\partial h_{2}$. We will not give the details of this limit here since they  
are   analogous to the ones presented above.  The dilaton goes to zero, the 
coordinate $\psi$ and $S_2^2$ produce a non-trivial $S^3$ and there is 
non-vanishing   RR 3-form flux through this three sphere. 
Hence the interpretation of this singular  solution is given by a probe D5-brane with 
worldvolume $S_1^2\times AdS_{4}$.

\newpage

\section{The AdS/CFT dual interface gauge theory}\label{holdual}
\setcounter{equation}{0}

The holographic  interpretation 
\cite{Maldacena:1997re, Gubser:1998bc, Witten:1998qj}
(for reviews, see \cite{D'Hoker:2002aw}, \cite{Aharony:1999ti})
of the original Janus solution  as  well as the $\cN=1$  and  $\cN=4$  supersymmetric generalizations are given by  interface 
conformal field theories respectively in  \cite{Bak:2003jk}, \cite{D'Hoker:2006uu}, and \cite{degAdS4}. The dual four-dimensional field theory  lives on two 
four-dimensional half-spaces glued together at a three-dimensional interface.
\begin{figure}[tbph]
\begin{center}
\epsfxsize=3.0in
\epsfysize=3.6in
\epsffile{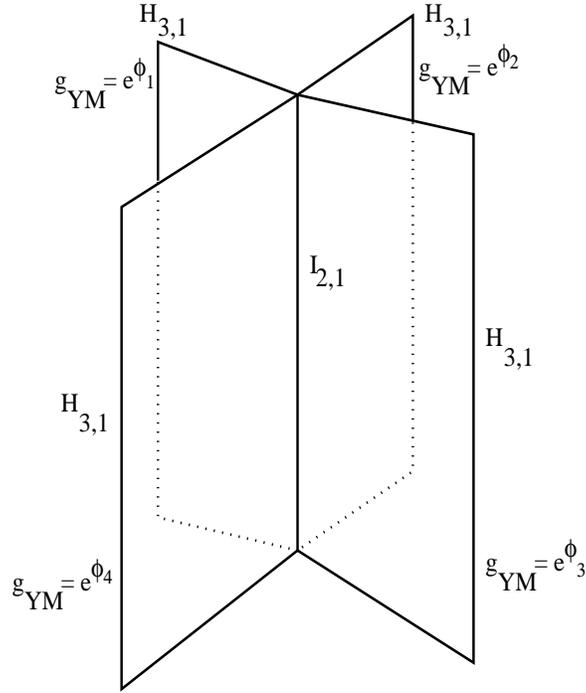}
\label{figure7}
\caption{Dual interface theory for the genus 1 multi-Janus solution}
\end{center}
\end{figure}
The boundary structure is found by determining where the various components of 
the  metric blow up. Using the Poincar\'e patch metric for the $AdS_{4}$ metric 
the ten-dimensional metric is given by
\be
ds^{2}= 4\rho^{2}|d w |^2 + f_{4}^{2 }
\left( {-dt^{2}+dx_{1}^{2} +dx_{2}^{2}+dz^{2}\over z^{2}}\right)  
+ f_{1}^{2}ds_{S_1^2  }+ f_{2}^{2} ds_{S_2^2 }
\ee
It was shown in section \ref{secfour} that the metric is completely regular for 
appropriate choices of the parameters of the genus $g$ hyperelliptic Ansatz. 
The only place where a metric factor blows up is at the branch points 
$e_{i}$, for $i=1,2,\cdots , 2g-1$ (and infinity), where $f_{4}$ diverges.  
There is an additional divergence of the $AdS_{4}$ metric when $z\to 0$. 
The boundary structure becomes clearer if
one rewrites the metric as follows
\be
ds^{2}={f_{4}^{2} \over z^{2}}\left\{    -dt^{2}+dx_{1}^{2} +dx_{2}^{2}+dz^{2}   
+{z^{2 } \over f_{4}} \Big( 4 \rho^{2}|d u |^2 + f_{1}^{2}ds_{S_1^2 }
+ f_{2}^{2}ds_{S_2^2 } \Big) \right\}
\ee
Near the $2g+2$ branch points, $f_{4}$ diverges and there are $2g+2$
different  $3+1$-dimensional half-spaces $H_{1,3}^{(i)},\; i=1,2,\cdots 2g+2$ 
each spanned by $t,x_{1},x_{2},z$. Each one of these spaces has a boundary 
at $z=0$. Since at $z=0$ 
the overall metric factor blows up, too,  there is an additional boundary component 
given by a $2+1$-dimensional space $I_{1,2}$ spanned by $t,x_{1},x_{2}$.  

\smallskip

The behavior of the fields near the boundary was worked out in subsection 
\ref{asymbranch}. The standard holographic relations  \cite{Witten:1998qj} 
allow the identification of the dual conformal field theory. The detailed analysis  
 is the same as the one  presented in \cite{degAdS4} and will not be repeated here. 
 The result that the holographic theory is an interface theory defined by $2g+2$ 
 copies of $\cN=4$ SYM living on (independent) $3+1$-dimensional half-spaces 
 $H_{1,3}^{(i)}$. The boundaries  of the half-spaces are all connected via a 
 $2+1$-dimensional interface $I_{1,2}$.

\smallskip

The SYM theories on the $i$-th half-space (which is associated with the branch 
point $e_{i}$), has the coupling constant which can be determined from (\ref{asymdil}). 
\be\label{symcoupl}
g_{\;YM}^{{(i)}}=\left | {Q_2(e_i) \over Q_1(e_i)}\right| 
\ee
In figure 10, we sketch the situation for the  $g=1$ case which has four 
boundary components. Note that in  figure 10, the four-dimensional spaces 
are embedded in a higher-dimensional space. This is mainly done for clarity, 
but is also  natural from a holographic perspective.  A different interpretation 
uses  a ``folding trick" employed in  \cite{Bachas:2001vj} where  $2g+2$ 
copies of $\cN=4$  SYM fields with coupling constants  (\ref{symcoupl}) exist independently on a  single $3+1$-dimensional half-space and only interact 
on the $2+1$-dimensional interface.

\newpage

\section{Discussion}
\setcounter{equation}{0}

In this paper, we have found an infinite family of non-singular solutions 
to the half-BPS  equations and Type IIB supergravity field equations on 
$AdS_4 \times S^2\times S^2\times \Sigma$ 
with $SO(2,3) \times SO(3) \times SO(3)$
symmetry. The general local solution was constructed  in the companion 
paper \cite{degAdS4}, and is specified  by two harmonic functions $h_1$ 
and $h_2$ on a Riemann surface $\Sigma$ with boundary. The class of 
non-singular solutions found in this paper is parameterized by a genus 
$g$ hyper-elliptic surface $\Sigma$ with boundary.
The branch points of $\Sigma$ are restricted to lie on the real axis.  
Global regularity is obtained by enforcing the  conditions 
(R1-R7) of subsection 3.5 on the harmonic functions $h_1$ and $h_2$. 
The ordering (\ref{totalorder}) of the $2g+2$ branch points and $g+1$ branch 
cuts  on the real line is a one-dimensional analog of the coloring of the 
two-dimensional plane in the ``bubbling  AdS" solutions\footnote{Similar observations were made in \cite{Gomis:2006cu}.}  of \cite{Lin:2004nb}. 
The conditions (R1-R7) in addition require that the real 
zeros of the differentials $\p h_1$ and $\p h_2$ obey a relative order 
amongst themselves and with  the branch points.

\smallskip

There are several directions for further research concerning the half-BPS 
interface theories. Our regular solutions were found amongst the general local 
solution of \cite{degAdS4} by imposing regularity and topology conditions. 
The topology conditions we enforced guarantee that there are asymptotic 
regions, where the geometry approaches $AdS_5\times S^5$. It would be 
interesting to investigate whether these conditions are necessary for
regularity. In particular, is it possible to find solutions on Riemann surfaces 
with topology  different from the ones  we have used ? For example, 
can one find regular solutions with  $h_1, h_2>0$ inside $\Sigma$, but 
with $W>0$ as well ? What is their topology ?

\smallskip

It would also be interesting to analyze the proposed dual interface CFT of 
the multi-Janus solutions in more detail; in particular how does one picture 
the $2g+2$ half-spaces glued together over a three-dimensional interface, 
and what are the properties of the corresponding interface super 
Yang-Mills theory ? For example, one feature of our construction is that 
the number of half-spaces  always has to be even. 
It is an interesting question whether this is required for the preservation of 
supersymmetry on the interface along the lines of the arguments of \cite{degsusy},
or is already required in order to have a consistent Dirac equation for the fermion
fields in the problem.

\smallskip

A new feature of the multi-Janus solutions (which has genus $g \geq 1$,
while the supersymmetric Janus solution had genus $g=0$) is the presence of
topologically non-contractible 3-cycles, and associated non-vanishing 
NSNS and RR 3-form charges. This fact, and the behavior of our solution in 
the limit where a branch cut collapses to a point, as was discussed in subsection 
\ref{problim}, shows that the solutions found in this paper are the fully 
back-reacted versions corresponding to the insertion of probe D5 and NS5 branes 
in $AdS_5 \times S^5$ considered in \cite{Karch:2000gx,DeWolfe:2001pq,Erdmenger:2002ex}.
Therefore, our half-BPS interface solutions  provide 
a new example of ``delocalization". The defect conformal field theories 
associated with the probe 5-branes contain extra degrees of freedom 
coming from the strings localized at the intersection of the three and five 
branes. In the fully back-reacted geometry, the localized probe branes are 
replaced by a full non-singular geometry with flux, and the extra degrees 
of freedom, supported only on the defect, have disappeared. 
It would be interesting to investigate whether this 
phenomenon is connected to the observations of 
\cite{Marolf:1999uq,Gomberoff:1999ps}.

\smallskip

A closely related class of supergravity solutions  has  
$SO(2,1)\times SO(3)\times SO(5)$ 
symmetry and is described by an $AdS_2 \times S^2 \times S^4 \times \Sigma $ 
Ansatz.  The gravitational solution is interpreted as the fully back reacted 
 geometry dual to  half-BPS Wilson loops \cite{Lunin:2006xr,Gomis:2006sb,Yamaguchi:2006te}. A detailed analysis of this solution  applying  the methods of this paper can be found in a further companion paper \cite{EDJEMG3}.

\bigskip

\bigskip

\noindent
{\large \bf Acknowledgements}

\bigskip

We are happy to acknowledge a useful conversation with George Morales
on planar electrostatics problems. 

\smallskip

This work 
was supported in part by a National Science Foundation (NSF) Physics Division 
grant  PHY-04-56200.

\newpage

\appendix

\section{No Regular solutions with complex poles and cuts}
\setcounter{equation}{0}

While it would seem natural to consider solutions in which poles and branch cuts
appear on the inside of the Riemann surface $\Sigma$ (as opposed to the 
branch cuts and poles on the boundary, which were explored in the main body of this
paper), we shall show in this appendix that no such regular solutions exist.
We begin by working out the example with a single interior pole.

\subsection{Janus plus one complex pole}

We work in the lower half-plane with the variable $u$ such that $\Im(u)<0$.
Including a complex pole\footnote{The square $p^2$ is used here for later 
convenience; without loss of generality, we choose $\Re(p)<0< \Im (p)$.}
 $p^2$, with $\Im (p^2) <0$ requires that we 
also include the complex pole $\bar p^2$. To keep the behavior at $u =\infty$
unchanged, we need to also include a complex zero $c$ and its complex conjugate $\bar c$.
Combining all these ingredients, we have the following complex pole 
generalization of the Janus solution,
\bea
\p h_1 & = & -i {(u-\a) (u - c) (u - \bar c) \over (u- p^2) (u - \bar p ^2) \sqrt{u}^3} du
\no \\
\p h_2 & = & - {(u-\b) (u - c) (u - \bar c) \over (u- p^2) (u - \bar p ^2) \sqrt{u}^3} du
\eea
Clearly, when $c=p^2$, the Janus solution is obtained as a limiting case.
Negativity of $W$ on the lower half $u$-plane, and positivity at the poles $u=0,\infty$  
requires that $\b >0$ and $\a <0$, as in the Janus case. Next, we impose 
vanishing Dirichlet boundary conditions and positivity of $h_1$ and $h_2$
throughout the lower half-plane, for which we need the functions $h_1$ and $h_2$.

\smallskip

To uniformize the square root, we change variables to $w^2 = u$ with 
$\Re(w)<0$, and $\Im (w) >0$, as in (\ref{domain}), so that $du/\sqrt{u}^3= 2 dw/w^2$, 
and   decompose into elementary fractions,
\bea
{(w^2 -\a) (w^2  - c) (w^2 - \bar c) \over w^2 (w^2 - p^2) (w^2 - \bar p ^2) }
& = & 
1+{A_0 \over w^2} + {A_1 \over w^2 - p^2 } + { \bar A_1 \over w^2 - \bar p^2}
\no \\
{(w^2 -\b) (w^2  - c) (w^2 - \bar c) \over w^2 (w^2 - p^2) (w^2 - \bar p ^2) }
& = & 
1+{B_0 \over w^2} + {B_1 \over w^2 - p^2 } + { \bar B_1 \over w^2 - \bar p^2}
\eea
where
\bea
\label{periods1ap}
A_0 = - \a {|c|^2 \over |p|^4} >0 & \hskip 1in & 
A_1 = {(p^2 -\a) (p^2  - c) (p^2 - \bar c) \over p^2  (p^2 - \bar p ^2) }
\no \\
B_0 = - \b {|c|^2 \over |p|^4} <0
& \hskip 1in & 
B_1 = {(p^2 -\b) (p^2  - c) (p^2 - \bar c) \over p^2  (p^2 - \bar p ^2) }
\eea
The functions $h_1$ and $h_2$ may then be obtained by elementary integrals. 
Decomposing the ratios $A_1/p$ and $B_1/p$ into their real and imaginary parts,
we obtain, 
\bea
h_1 & = & h_1 ^{(0)}- 2i(w - \bar w) + 2 i A_0 {\bar w - w\over |w|^2} 
+ \Re \left ( - i {A_1 \over p} \right ) 
 	\ln \left | {(w-p)( w + \bar p) \over (w+p)( w - \bar p)} \right |^2
\no \\ && \hskip 1in 
+ i \, \Im \left ( {-i  A_1 \over p} \right ) 
	\ln {(w-p) (\bar w + p) (\bar w+ \bar p)( w - \bar p) \over (\bar w- \bar p)( w + \bar p)
		(w+p) (\bar w -p)}
\no \\
h_2 & = & h_2 ^{(0)} - 2(w + \bar w)+ 2  B_0 {\bar w + w\over |w|^2} 
-  \Re \left ( {B_1 \over p} \right ) \ln \left | {(w-p)( w - \bar p) \over (w+p)(w + \bar p)} \right |^2
\no \\ && \hskip 1in 
- i \, \Im \left ( { B_1 \over p} \right )
	\ln {(w-p) (\bar w-  p)( \bar w + \bar p)(w+\bar p) \over (\bar w- \bar p)( w - \bar p)
	(w+p)  (\bar w +p)}
\eea
The logarithms in the last term in each expression are multiple-valued in $w$ 
around the pole $p$. Single-valuedness of $h_1$ and $h_2$ requires that 
$\Re(A_1/p)= \Im (B_1/p)=0$.
These conditions are equivalent to requiring vanishing periods 
of both differentials $\p h_1$ and $\p h_2$ around the pole $p$.
We parametrize the solutions to these equations explicitly by
\bea
\label{periodsab2}
A_1 = - i p a_1
\hskip 1in 
B_1 =   p b_1 \hskip 1in a_1, b_1 \in {\bf R}
\eea
Requiring $h_1$ to vanish for real $w<0$, and $h_2$ to vanish for 
imaginary $w$, we find that we must have $h_1^{(0)}=h_2^{(0)}=0$.
We then have the following simplified expressions, 
\bea
h_1 & = &-2i(w - \bar w)+2 i A_0 {\bar w - w\over |w|^2} 
- a_1 \ln \left | {(w-p)( w + \bar p) \over (w+p)( w - \bar p)} \right |^2
\no \\
h_2 & = &  -2(w + \bar w)+2  B_0 {\bar w + w\over |w|^2} 
-b_1  \ln \left | {(w-p)( w - \bar p) \over (w+p)(w + \bar p)} \right |^2
\eea
Since we have assumed that both $w$ and $p$ lie in the second quadrant,
\bea
\Re (w) <0 && \Re (p) <0
\no \\
\Im (w) >0 && \Im (p) >0
\eea
we have the inequalities,
\bea 
\label{ineq6}
 |  w-p | < | w - \bar p | & \hskip 0.5in &
|  w + \bar p | < | w + p  | 
\no \\
| w-p | < | w + \bar p  |  & \hskip 0.5in &
| w - \bar p | < | w + p  | 
\eea
Thus, positivity of $h_1$ and $h_2$ will be assured for $w$ in the 
second quadrant provided,
\bea
a_1, b_1 \geq 0
\eea
It  remains to solve (\ref{periods1ap}) and the period relations 
(\ref{periodsab2}). Combining both relations, we have, 
\bea
(p^2 - \a) (p^2 - c) (p^2-\bar c) & = & - i a_1 p^3 (p^2 - \bar p^2)
\no \\
(p^2 - \b) (p^2 - c) (p^2-\bar c) & = &  + b_1 p^3 (p^2 - \bar p^2)
\eea
We consider these two complex equations for $a_1$ and $b_1$  given,
and view $p^2$ and the position of the zero $c$ as determined by $a_1$, $b_1$,  
and the above period relations. Taking the ratio of the two equations, 
we obtain an equation for $p^2$,
\bea
p^2 = { \a b_1 + i \b \a_1  \over b_1 + i a_1}
\eea
The imaginary part of $p^2$ is given by
\bea
\label{Imp}
 \Im (p^2) =  {a_1 b_1 (\b - \a ) \over | a_1 - i b_1|^2} 
\eea
Under the conditions $a_1, b_1>0$, and using the inequality on the real
zeros $\beta - \alpha >0$, the pole $p^2$ actually lies in the {\sl upper} half-plane,
contrarily to our assumptions. 

\smallskip

Note that if we had taken $p^2$ in the 
upper half-plane instead, and $p$ in the first quadrant, the inequalities 
in the first line of (\ref{ineq6}) would remain unchanged, but the 
inequalities in the last line would get reversed, with the effect of 
now requiring $a_1\geq 0$ but $b_1 \leq 0$. From (\ref{Imp}), 
we would then find that now $p^2$ must be in the lower half-plane, 
which is again in contradiction with the assumptions made.
Thus, the formal solution with one complex pole can never satisfy all
the necessary conditions of regularity.

\subsection{The addition of general complex poles and branch cuts}

The arguments presented in the preceding subsection are essentially local
on the surface $\Sigma$. Thus, by analogous reasoning, the presence of
multiple complex poles is also ruled out. The arguments may also be 
generalized to complex branch cuts.


\newpage







\begin{thebibliography}{99}

{\small

\bibitem{degAdS4}
E. D'Hoker, John Estes, and M. Gutperle, 
``Exact Half-BPS type IIB Interface solutions I: 
Local solution and supersymmetric Janus",
  [arXiv:0705.0022].

\bibitem{Gomis:2006cu}
  J.~Gomis and C.~Romelsberger,
  ``Bubbling defect CFT's,''
  JHEP {\bf 0608}, 050 (2006)
  [arXiv:hep-th/0604155].

\bibitem{degsusy}
  E.~D'Hoker, J.~Estes and M.~Gutperle,
  ``Interface Yang-Mills, supersymmetry, and Janus,''
  Nucl. Phys. B {\bf 753} (2006) 16,
  [arXiv:hep-th/0603013].

\bibitem{DeWolfe:2001pq}
  O.~DeWolfe, D.~Z.~Freedman and H.~Ooguri,
   ``Holography and defect conformal field theories,''
  %
  Phys.\ Rev.\ D {\bf 66}, 025009 (2002)
  [arXiv:hep-th/0111135].

\bibitem{Lin:2004nb}
  H.~Lin, O.~Lunin and J.~M.~Maldacena,
  ``Bubbling AdS space and 1/2 BPS geometries,''
  JHEP {\bf 0410}, 025 (2004)
  [arXiv:hep-th/0409174].

  


\bibitem{Lunin:2006xr}
  O.~Lunin,
  ``On gravitational description of Wilson lines,''
  JHEP {\bf 0606} (2006) 026
  [arXiv:hep-th/0604133].
  
\bibitem{Gomis:2006sb}
  J.~Gomis and F.~Passerini,
  ``Holographic Wilson loops,''
  JHEP {\bf 0608} (2006) 074
  [arXiv:hep-th/0604007].


\bibitem{Yamaguchi:2006te}
  S.~Yamaguchi,
  ``Bubbling geometries for half BPS Wilson lines,''
  arXiv:hep-th/0601089.
  
 \bibitem{EDJEMG3} 
   E.~D'Hoker, J.~Estes and M.~Gutperle, ``Gravity duals of half-BPS Wilson loops", (2007).
 
  

\bibitem{Schwarz:1983qr}
  J.~H.~Schwarz,
  ``Covariant Field Equations Of Chiral N=2 D = 10 Supergravity,''
  Nucl.\ Phys.\ B {\bf 226} (1983) 269.

\bibitem{Howe:1983sr}
  P.~S.~Howe and P.~C.~West,
  ``The Complete N=2, D = 10 Supergravity,''
  Nucl.\ Phys.\ B {\bf 238} (1984) 181.



\bibitem{bateman}
A. Erdelyi, Editor, {\sl Higher transcendental Functions, Bateman Manuscript Project}, Vol II,
Chapter XIII, Robert E. Krieger Publishing Company, (1981).

\bibitem{magnus}
F. Oberhettinger and W. Magnus, {\sl Anwendung der elliptischen Funcktionen in
Physik un Technik}, Springer-Verlag (1948).

\bibitem{mumford}
D. Mumford, {\sl Tata Lectures on Theta}, Vol 1, (1982) Birkh\"auser.

\bibitem{DPIV}
E. D'Hoker and D.H. Phong, 
``Two-loop superstrings IV, The cosmological constant and modular forms",
Nucl. Phys. B {\bf 639} (2002) 129
[arXiv:hep-th/0111040].

\bibitem{Fay}
J. Fay, {\sl Theta Functiosn on Riemann surfaces}, Springer 1973;\\
E. D'Hoker and D.H. Phong, ``The geometry of string perturbation theory",
Rev. Mod. Phys. {\bf 60} (1988) 917.

\bibitem{Maldacena:1997re}
  J.~M.~Maldacena,
   ``The large N limit of superconformal field theories and supergravity,''
  %
  Adv.\ Theor.\ Math.\ Phys.\  {\bf 2}, 231 (1998)
  [Int.\ J.\ Theor.\ Phys.\  {\bf 38}, 1113 (1999)]
  [arXiv:hep-th/9711200].

\bibitem{Gubser:1998bc}
  S.~S.~Gubser, I.~R.~Klebanov and A.~M.~Polyakov,
   ``Gauge theory correlators from non-critical string theory,''
  %
  Phys.\ Lett.\ B {\bf 428}, 105 (1998)
  [arXiv:hep-th/9802109].

\bibitem{Witten:1998qj}
  E.~Witten,
   ``Anti-de Sitter space and holography,''
  %
  Adv.\ Theor.\ Math.\ Phys.\  {\bf 2}, 253 (1998)
  [arXiv:hep-th/9802150].

\bibitem{D'Hoker:2002aw}
  E.~D'Hoker and D.~Z.~Freedman,
  ``Supersymmetric gauge theories and the AdS/CFT correspondence,''
   in {\sl Strings, Branes, and Extra Dimensions},
   S.S. Gubser, J.D. Lykken, Eds,
   World Scientific (2004),
  arXiv:hep-th/0201253.

\bibitem{Aharony:1999ti}
  O.~Aharony, S.~S.~Gubser, J.~M.~Maldacena, H.~Ooguri and Y.~Oz,
  ``Large N field theories, string theory and gravity,''
  Phys.\ Rept.\  {\bf 323} (2000) 183
  [arXiv:hep-th/9905111].


 
\bibitem{Bak:2003jk}
  D.~Bak, M.~Gutperle and S.~Hirano,
   ``A dilatonic deformation of AdS(5) and its field theory dual,''
  %
  JHEP {\bf 0305}, 072 (2003)
  [arXiv:hep-th/0304129].


\bibitem{D'Hoker:2006uu}
  E.~D'Hoker, J.~Estes and M.~Gutperle,
  ``Ten-dimensional supersymmetric Janus solutions,''
  Nucl.\ Phys.\  B {\bf 757} (2006) 79
  [arXiv:hep-th/0603012].
  
\bibitem{Bachas:2001vj}
  C.~Bachas, J.~de Boer, R.~Dijkgraaf and H.~Ooguri,
  ``Permeable conformal walls and holography,''
  JHEP {\bf 0206} (2002) 027
  [arXiv:hep-th/0111210].





\bibitem{Karch:2000gx}
  A.~Karch and L.~Randall,
  ``Open and closed string interpretation of SUSY CFT's on branes with
  boundaries,''
  JHEP {\bf 0106} (2001) 063
  [arXiv:hep-th/0105132].



\bibitem{Erdmenger:2002ex}
  J.~Erdmenger, Z.~Guralnik and I.~Kirsch,
  ``Four-dimensional superconformal theories with interacting boundaries or
  Phys.\ Rev.\  D {\bf 66} (2002) 025020
  [arXiv:hep-th/0203020].


\bibitem{Marolf:1999uq}
  D.~Marolf and A.~W.~Peet,
  ``Brane baldness vs. superselection sectors,''
  Phys.\ Rev.\  D {\bf 60} (1999) 105007
  [arXiv:hep-th/9903213].


\bibitem{Gomberoff:1999ps}
  A.~Gomberoff, D.~Kastor, D.~Marolf and J.~H.~Traschen,
  ``Fully localized brane intersections: The plot thickens,''
  Phys.\ Rev.\  D {\bf 61} (2000) 024012
  [arXiv:hep-th/9905094].

}
 
 
\end{thebibliography}
\end{document}